\documentclass[10pt,canadian,english,journal=jpclcd,manuscript=letter,layout=onecolumn]{achemso}
\usepackage[T1]{fontenc}
\usepackage[utf8]{luainputenc}
\usepackage{color}
\usepackage{babel}
\usepackage{array}
\usepackage{multirow}
\usepackage{amsmath}
\usepackage{amssymb}
\usepackage{graphicx}
\usepackage[pdfusetitle,
 bookmarks=true,bookmarksnumbered=false,bookmarksopen=false,
 breaklinks=false,pdfborder={0 0 0},pdfborderstyle={},backref=false,colorlinks=true]
 {hyperref}
\usepackage{pdfpages}

\graphicspath{{./Figures/}{./}}

\makeatletter

\title{Separating orders of response in transient absorption and coherent
multi-dimensional spectroscopy by intensity variation}

\author{Jacob J. Krich}

\affiliation{Department of Physics, University of Ottawa, Ottawa, ON K1N 6N5,
Canada}

\alsoaffiliation{Nexus for Quantum Technologies, University of Ottawa, Ottawa, ON
K1N 6N5, Canada}

\altaffiliation{Contributed equally to this work}

\email{jkrich@uottawa.ca}

\author{Luisa Brenneis}

\affiliation{Institut für Physikalische und Theoretische Chemie, Universität Würzburg,
Am Hubland, 97074 Würzburg, Germany}

\altaffiliation{Contributed equally to this work}

\author{Peter A. Rose}

\affiliation{Department of Physics, University of Ottawa, Ottawa, ON K1N 6N5,
Canada}

\author{Katja Mayershofer}

\affiliation{Institut für Physikalische und Theoretische Chemie, Universität Würzburg,
Am Hubland, 97074 Würzburg, Germany}

\author{Simon Büttner}

\affiliation{Institut für Physikalische und Theoretische Chemie, Universität Würzburg,
Am Hubland, 97074 Würzburg, Germany}

\author{Julian Lüttig}

\affiliation{Department of Physics, University of Michigan, 450 Church St., Ann
Arbor, MI 48109, USA}

\author{Pavel Mal\'{y}}

\affiliation{Faculty of Mathematics and Physics, Charles University, Ke Karlovu
5, 121 16 Prague, Czech Republic}

\author{Tobias Brixner}

\affiliation{Institut für Physikalische und Theoretische Chemie, Universität Würzburg,
Am Hubland, 97074 Würzburg, Germany}

\alsoaffiliation{Center for Nanosystems Chemistry (CNC), Universität Würzburg, Theodor-Boveri-Weg,
97074 Würzburg, Germany}

\email{tobias.brixner@uni-wuerzburg.de}

\providecommand{\tabularnewline}{\\}

\RequirePackage{fix-cm}
\usepackage{braket}
\usepackage{xr}

\usepackage[fontsize=9pt]{fontsize}

\usepackage[table]{xcolor}

\usepackage{hyperref} 
\hypersetup{ colorlinks, 
linkcolor={red!50!black}, 
citecolor={blue!80!black}, 
urlcolor={blue!80!black} }

\makeatother

\begin{document}
\selectlanguage{canadian}%
\providecommand{\comment}[1]{\textbf{[#1]}}

\global\long\def\braket#1#2{\Braket{#1|#2}}%

\global\long\def\bra#1{\Bra{#1}}%

\global\long\def\ket#1{\Ket{#1}}%

\global\long\def\cvec#1{\boldsymbol{#1}}%

\global\long\def\Smax{S_{\text{max}}}%

\global\long\def\Isat{I_{\text{sat}}}%

\global\long\def\Isattwo{I_{\text{sat}}}%

\global\long\def\nQ{n\text{\text{Q}}}%

\global\long\def\rQ{r\text{\text{Q}}}%

\global\long\def\I{\mathcal{I}}%

\global\long\def\Sprime{S^{\prime}}%

\global\long\def\Imax{I_{\text{max}}}%
\begin{tocentry}
\includegraphics[width=3.25in]{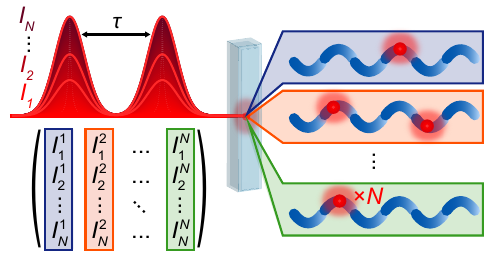}
\end{tocentry}
\begin{abstract}
Interpretation of time-resolved spectroscopies such as transient
absorption (TA) or two-dimensional (2D) spectroscopy often relies
on the perturbative description of light-matter interaction. In many
cases the third order of nonlinear response is the leading and desired
term. When pulse amplitudes are high, higher orders of light-matter
interaction can both distort lineshapes and dynamics and provide valuable
information. Here, we present a general procedure to separately measure
the nonlinear response orders in both TA and 2D spectroscopies, using
linear combinations of intensity-dependent spectra. We analyze the
residual contamination and random errors and show how to choose optimal
intensities to minimize the total error in the extracted orders. For
an experimental demonstration, we separate the nonlinear orders in
the 2D electronic spectroscopy of squaraine polymers up to 11\textsuperscript{th}
order.
\end{abstract}

Nonlinear spectroscopies -- from transient absorption (TA) to coherent
multidimensional spectroscopies -- require a careful balance when
choosing the intensities of each pulse. In general, increasing pulse
intensity improves the signal-to-noise ratio (SNR). The measured signal
is commonly described as a power series in the amplitudes of the pulses\cite{MukamelBook}
and spectroscopies are most commonly interpreted using the lowest-order
term in that series. However, for larger pulse intensities the neglected
higher-order terms provide a contamination of the desired signals,
which physically represents extra interactions with the pulses. Such
higher-order contamination distorts the signal spectral shape and
dynamics and complicates the interpretation.

\begin{figure}
\includegraphics[width=3.375in]{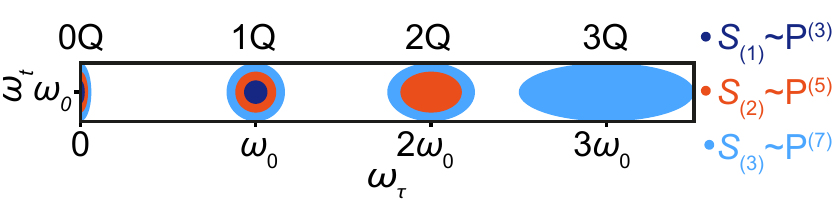}
\caption{Schematic higher-order contributions in 2D spectra with a weak probe
showing $\protect\nQ$ signals at multiples of the carrier frequency
$\omega_{0}$, using excitation frequency $\omega_{\tau}$ and detection
frequency $\omega_{t}$. The lowest-order contribution to each $\protect\nQ$
signal is $S_{(2n)}$, which is $2n^{\text{th}}$ order in the pump
amplitudes and order $2n+1$ in the total field amplitude. The elongation
along $\omega_{\tau}$ of the $\protect\nQ$ signal denotes the spectral
width of the $n^{\textrm{th}}$ harmonic of the excitation spectrum.
For each $\protect\nQ$ signal, there are several contributions of
different orders. While only $\protect\nQ$ signals with $\omega_{\tau}\protect\geq0$
are shown, equivalent $\protect\nQ$ signals exist for $\omega_{\tau}<0$.
\label{fig:nQ_schematic-1}}
\end{figure}

We recently showed that these orders of response can be separated
in TA spectroscopy\cite{Maly23} and in excitation-frequency integrated
2D spectroscopy\cite{Luttig23} using a procedure called ``intensity
cycling,'' enabling the extraction of third-order spectra even when
using pulses intense enough to produce contaminated spectra. Furthermore,
contributions from higher nonlinear orders can be separated and extracted,
which reveal valuable information about multiply excited states and
processes such as exciton--exciton annihilation.\cite{Maly23,Rose23}
In that intensity-cycling scheme, spectra are collected using $N$
pump intensities $I_{p}$ obeying the intensity cycling ratios, $I_{p}=4I_{0}\cos^{2}(\pi p/2N)$
for $p=0,\dots,N-1$, where $I_{0}$ is a base intensity. The first
$N$ response orders are extracted using formulae derived from the
connection between TA spectra and $n$-quantum ($\nQ$) signals that
are frequently studied in two-dimensional electronic spectroscopy
(2DES).\cite{Maly23,Luttig23,Heshmatpour20,Luttig23a} The $n$Q signals
appear in the $-n\vec{k}_{1}+n\vec{k}_{2}+\vec{k}_{3}$ direction
in a phase-matched 2DES experiment, and are centered at $\omega_{\tau}=n\omega_{0}$
where $\omega_{0}$ is the carrier frequency of the pump pulses (Figure
\ref{fig:nQ_schematic-1}), $\omega_{\tau}$ is the excitation frequency,
and $\vec{k}_{i}$ are the wavevectors of the optical pulses. Such
$\nQ$ signals have been used to access higher-order responses\cite{ding_heterodyned_2005,yu_observation_2018,Maly20,Heshmatpour20,bruggemann_nonperturbative_2011,Turner10},
beyond third order, as the leading order response increases with each
multiple of $\omega_{0}$. Intensity cycling also gives access to
such higher-order signals using simpler TA measurements. Intensity
cycling is in some aspects even superior to the $\nQ$ measurements
since, as visualized in Figure~\ref{fig:nQ_schematic-1}, each $\nQ$
signal has contributions from many response orders, not only the leading
$(2n+1)^{\text{th}}$-order contribution; intensity cycling separates
those contributions.

Here, we generalize from TA to coherently detected 2DES.\cite{Brixner05,oliver_correlating_2014,Turner11,Fuller15,Dostal18}
In this generalization, data collected with $N$ pump intensities
allows one to infer the first $N$ response orders. We show three
major conclusions from this generalization. First, that the extraction
of higher-order signals does not require the specific pump intensity
ratios from the previously published intensity-cycling protocol but
is feasible with arbitrary intensities. This generalization permits
the re-analysis of previously collected and already published intensity-dependent
data. Second, the nonlinear order separation is not limited to TA
as in the prior scheme \cite{Maly23} but can also be applied to 2DES.
Third, in the case where the pump pulses are identical except in arrival
time, we quantify how random experimental noise and remaining contamination
from unextracted higher orders affect the inferred response orders
and use that theory to choose pump intensities that minimize the total
error.

We demonstrate the technique experimentally on a squaraine copolymer
{[}SQA-SQB{]}\textsubscript{18\textsubscript{}} (molecular structure
shown in Supporting Information Figure~S1), with an average of
18 dimer units,\cite{Voelker2014,Maly20} and show the extraction
up to the 11$^{\text{th}}$ order of response. Finally, we present
self-consistency conditions to verify that the response orders have
been correctly extracted.

We begin by defining the response orders and showing how they can
be extracted. We consider a sequence of $L=2$ (TA) or $L=3$ (2DES)
pulses which interact with the system. The electric field at time
$t$ and position $\vec{r}$ can be written in the vicinity of the
sample as
\begin{equation}
\vec{E}(t,\vec{r})=\sum_{l=1}^{L}\vec{e}_{l}\varepsilon_{l}\left(t-\frac{\vec{k}_{l}\cdot\vec{r}}{\omega_{l}}-t_{l}\right)e^{-i(\omega_{l}(t-t_{l})-\vec{k}_{l}\cdot\vec{r})}+\text{c.c.},\label{eq:E}
\end{equation}
where for pulse $l$, $\varepsilon_{l}(t)$ is the complex pulse envelope
with any chirp included, $\omega_{l}$ is the central frequency, $t_{l}$
is the arrival time, and $\vec{e}_{l}$ is the polarization, which
can itself vary in space or time. We now focus on the case of coherently
detected 2DES with $L=3$ pulses and later show that the conclusions
also apply to TA.

Consider that we vary the amplitude of the first two pulses, multiplying
both $\varepsilon_{1}$ and $\varepsilon_{2}$ by a factor $\lambda>0$
but leaving $\varepsilon_{3}$ unchanged. We do not need $\varepsilon_{1}(t)$
and $\varepsilon_{2}(t)$ to be identical. We further define a dimensionless
intensity $I=\lambda^{2}$. We first write down the conclusion of
how the signal scales with $I$ and introduce a new notation $S_{(j)}$
to show the response orders in $\lambda^{j}$; we then motivate the form.
In the absence of noise, the nonlinear signal can be written as
\begin{equation}
S(\tau,T,t,I)=\sum_{j=1}^{\infty}S_{(2j)}(\tau,T,t)I^{j},\label{eq:S_orders_in_I}
\end{equation}
where $\tau=t_{2}-t_{1}$ is the coherence time, $T=t_{3}-t_{2}$
is the population time, and $t$ is the signal time, and we suppress
the dependence of the signal on $\tau$, $T$, and $t$ for much of
this discussion. 
For phase-matched TA and 2DES, $S_{(j)}=0$ when $j$ is odd. 
Consider that we want to extract the first $N$ orders
$S_{(2)}$ to $S_{(2N)}$ in Eq.~\ref{eq:S_orders_in_I}. To this purpose, we define
\begin{equation}
	\Sprime(\tau,T,t,I)=\sum_{j=1}^{N}S_{(2j)}(\tau,T,t)I^{j},\label{eq:Sprime_orders_in_I}
\end{equation}
where $\Sprime(I)\approx S(I)$ if $I$ is small enough that all terms
$S_{(2j)}I^{j}$ are negligible for $j>N$. The orders $S_{(2j)}$ can
be extracted by determining $\Sprime(I)$ at $N$ different intensities
$I_{k}$ with $k=1,...,N$, from which Eq.~\ref{eq:Sprime_orders_in_I}
implies
\begin{align}
\left(\begin{array}{c}
\Sprime(I_{1})\\
\Sprime(I_{2})\\
\vdots\\
\Sprime(I_{N})
\end{array}\right)= & \left(\begin{array}{cccc}
I_{1} & I_{1}^{2} & \dots & I_{1}^{N}\\
I_{2} & I_{2}^{2} & \dots & I_{2}^{N}\\
\vdots &  & \ddots\\
I_{N} & I_{N}^{2} &  & I_{N}^{N}
\end{array}\right)\left(\begin{array}{c}
S_{(2)}\\
S_{(4)}\\
\vdots\\
S_{(2N)}
\end{array}\right).\label{eq:VdM}
\end{align}
We rewrite Eq.~\ref{eq:VdM} as $\Sprime(I_{k})=\sum_{n}V_{kn}S_{(2n)}$,
with $V_{kn}=I_{k}^{n}$. The matrix $V$ is close to being a Vandermonde
matrix, though it does not have the first column of ones in a conventional
Vandermonde matrix because in coherently detected 2DES the pump-independent
background is typically removed, resulting in $S_{(0)}=0$. The inverse
matrix $V^{-1}$ allows extracting $S_{(2n)}$ from the $\Sprime(I_{k})$
as $S_{(2n)}=\sum_{k}V_{nk}^{-1}\Sprime(I_{k})$. But the measured
data are $S(I_{k})$ and not $\Sprime(I_{k})$, and the same procedure
performed on the measured $S(I_{k})$ gives inferred values of the
$S_{(2n)}$, which we call $\hat{S}_{(2n)}$,
\begin{equation}
\hat{S}_{(2n)}\equiv\sum_{k=1}^{N}V_{nk}^{-1}S(I_{k}).\label{eq:VdM_inverted}
\end{equation}
When noise and contamination beyond the truncation of Eq.~\ref{eq:Sprime_orders_in_I}
are negligible, the extracted $\hat{S}_{(2n)}$ equal the underlying
$S_{(2n)}$. We show below how to estimate the contributions of both
sources of error to $\hat{S}_{(2n)}$ for intensities $I_{k}$.

We now motivate Eq.~\ref{eq:S_orders_in_I} and describe the physical
meaning of the expansion terms $S_{(n)}$ by connecting back to the
standard theory of nonlinear spectroscopy. In a coherently detected
measurement, the emitted electric field is determined by the induced
nonlinear polarization $P^{(\text{NL})}$ of the sample. For our derivation,
we only need that the detected fields are proportional to $P^{(NL)}(\omega_{t})$,
but the proportionality can depend on $\omega_{t}$, as long as it
is independent of the pump-pulse parameters. For the present analysis,
we use $P^{(NL)}$ as a proxy for the measured signal. The polarization
$P^{\text{(NL)}}=P^{(3)}+P^{(5)}+\dots$ can be written perturbatively
in $E(t)$, see Eq.~\ref{eq:E}, in terms of system response functions
$R^{(n)}$,\cite{MukamelBook} where for example
\begin{align}
P^{(3)}(t)= & \int dt_{3}dt_{2}dt_{1}E(t-t_{3})E(t-t_{3}-t_{2})E(t-t_{3}-t_{2}-t_{1})R^{(3)}(t_{3},t_{2},t_{1}),\label{eq:P3}\\
P^{(5)}(t)= & \int dt_{5}\dots dt_{1}E(t-t_{5})E(t-t_{5}-t_{4})E(\dots)E(\dots)E(\dots)R^{(5)}(t_{5},\dots,t_{1}),\label{eq:P5}
\end{align}
where we ignore the polarizations of the electric fields for simplicity.
We now consider 2DES in a pump-probe geometry,\cite{Shim09} for which
the phase-matching condition causes only odd orders to contribute
to the signal. The order-separation technique is most straightforward
in the weak-probe limit, where detected signals are proportional only
to the first power of the probe amplitude $\varepsilon_{3}$; we consider
that limit first and then generalize. When $E$ in Eqs.~\ref{eq:P3},
\ref{eq:P5} is expanded using Eq.~\ref{eq:E}, in every term that
contributes to the signal, one of the $E$ fields in Eq.~\ref{eq:P3}
or \ref{eq:P5} must be a contribution from the probe $\varepsilon_{3}(t)$,
and the other two or four $E$ fields must be contributions from the
pump pulses; combinations without $\varepsilon_{3}$ do not reach
the detector, and combinations with higher powers of $\varepsilon_{3}$
are neglected in the weak-probe limit. In all terms that contribute
to the signal, when $\varepsilon_{1}$ and $\varepsilon_{2}$ each
scale with $\lambda$, it is easy to see that $P^{(2n+1)}$ scales
with $\lambda^{2n}=I^{n}.$ By comparison with Eq.~\ref{eq:S_orders_in_I},
it follows that $S_{(2n)}I^{n}=P^{(2n+1)}$, so $S_{(2n)}$ reports
on $R^{(2n+1)}$. Extracting the orders $S_{(2n)}$ using Eq.~\ref{eq:VdM}
is then equivalent to extracting the standard orders of response $R^{(2n+1)}$.
In this notation, superscripts indicate orders of response in the
total electric field and subscripts indicate orders of response in
the electric field of the pulses whose amplitudes scale with $\lambda$ -- 
the pump pulses in our case.

\begin{figure}
\includegraphics{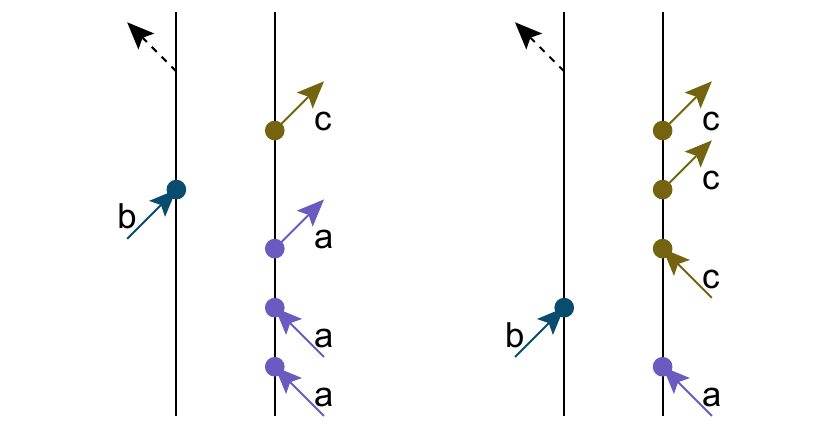} 
\caption{Two sample fifth-order pathways contributing to coherently detected
2D spectroscopy. While both pathways contribute to $R^{(5)}$, in
an experiment in which the pump pulse (a,b) amplitudes are varied
and the probe pulse (c) amplitude is unvaried, the left pathway contributes
to $S_{(4)}$, and the right pathway contributes to $S_{(2)}$, as
can be seen by counting the number of arrows containing either a or
b. The right pathway only contributes significantly if the probe pulse
is strong. \label{fig:diagrams}}
\end{figure}

If the probe pulse is strong enough that its contributions beyond
first order are not negligible, then the mapping from $S_{(n)}$ to
$R^{(m)}$ is less straightforward. In this case, $P^{(3)}$ is still
linear in $\varepsilon_{3}$, but $P^{(5)}$ can have contributions
that are linear or cubic in $\varepsilon_{3}$; quadratic contributions
in $\varepsilon_{3}$ would not obey the phase-matching condition.
We can break $P^{(5)}$ into portions that scale with $\lambda^{2}=I$
and portions that scale with $\lambda^{4}=I^{2}$. The former contributes
to $S_{(2)}$ and is cubic in $\varepsilon_{3}$ while the latter
contributes to $S_{(4)}$ and is linear in $\varepsilon_{1}$; example
pathways contributing to both are shown in Figure~\ref{fig:diagrams}.
Pathways with $n$ total pulse interactions contribute to $P^{(n)}$,
and pathways that have 2$n$ interactions with the pump pulses contribute
to $S_{(2n)}$. In general, $P^{(2n+1)}$ contains terms that contribute
to $S_{(2)},\dots,S_{(2n)}$. Conversely, $S_{(2n)}$ in principle has
contributions from $P^{(2n+1)}$ and higher orders, though the probe
must have high intensity for the higher orders to contribute significantly.
TA spectroscopy ($L=2$) with the pump intensity varied also obeys
Eq.~\ref{eq:S_orders_in_I} (without a $\tau$ dependence) and the
same connections between $S_{(n)}$ and $R^{(m)}$ as in 2DES.

In summary, $R^{(n)}$ is the usual $n^{\text{th}}$-order system
response function with $n$ being the order in the total applied electric
field including pump and probe pulses. When the amplitudes of the
pump pulses $\varepsilon_{1}$ and $\varepsilon_{2}$ are each scaled
by the same factor $\lambda$, $S_{(n)}$ is the $n^{\text{th}}$-order
response of the signal in $\lambda$. The
particular pulse shapes defined in Eq.~\ref{eq:E}, i.e., without
amplitude scaling ($\lambda=1$), affect both $P^{(n)}$ (see Eqs.~\ref{eq:P3},
\ref{eq:P5}) and $S_{(n)}$. Similarly, note that extractions performed
with Eq.~\ref{eq:VdM_inverted} depend only on the ratios of the
$I_{k}$; if all intensities $I_{k}$ are expressed as multiples of
a base intensity $I_{0}$, then $\hat{S}_{(2n)}$ extracted using Eq.~\ref{eq:VdM_inverted}
simply scales with $I_{0}^{n}$ as $I_{0}$ varies, which we discuss
further in Supporting Information Section S2.

Real optical pulses also have spatial profiles 
not described by Eq.~\ref{eq:E}, as in the usual case of focused pulses
with Gaussian lateral profiles or even with spatial chirp. 
In the usual limit of local response,\cite{MukamelBook} where molecules
each independently interact with the optical fields (e.g., see Eq.~5.13 in Ref.~\citenum{MukamelBook}), 
the extraction that we describe here applies equally well to optical pulses 
with spatial profiles. 
As long as the entire pulse amplitude is scaled
with the same factor $\lambda$, the extraction of response orders $S_{(2n)}$ 
is robust to spatial variation of the pulses
as every molecule experiences the same ratios of pulse amplitudes, so they each 
individually obey Eqs.~\ref{eq:P3},\ref{eq:P5}, and the scaling of these terms
with $\lambda^{2n}$ is the same for every molecule. 
The experimental work of Refs.~\citenum{Maly23,Luttig23} and this manuscript 
use probe pulses more tightly focused than the pump pulses, to ensure
that the pump intensities are approximately uniform over the studied molecules; 
while this choice is helpful for interpretation, the extraction of the orders 
of response of the signals does not require it. 
Nonlocal effects, such as cascading processes,\cite{Blank99} are beyond the scope of the
present work. We have previously found them not to be present in similar
samples of equivalent optical depth.\cite{Maly23}

The intensity cycling method of Ref.~\citenum{Maly23} reduces to
the form of Eq.~\ref{eq:VdM}, but Eq.~\ref{eq:VdM} is more general
because it shows that any sufficiently small intensities can work
rather than only the specific intensity cycling ratios reported earlier.
However, the new possibility of choosing arbitrary intensities raises
the question of what intensities are optimal for extracting the $S_{(n)}$.
Choosing intensities requires a balance between random and systematic
errors. Higher pump intensity generally increases the signal and improves
the SNR. But higher intensities also increase the systematic contamination
error from higher-order terms in the truncated series of Eq.~\ref{eq:S_orders_in_I}.
We demonstrate how to find the optimal intensities that balance systematic
and random error for TA and 2DES spectroscopies in the pump-probe
geometry with identical pump pulses in the weak-probe limit. Consider
that we are performing an $N$-intensity extraction of $N$ orders up to
$\hat{S}_{(2N)}$. Then we can write the measured signal $S(I)$ as
\begin{equation}
S(I)=\Sprime(I)+c(N,I)+\eta\label{eq:noContaminationS(I)}
\end{equation}
where $\Sprime(I)$ is the ideal signal in the absence of contamination
or random error in Eq.~\ref{eq:Sprime_orders_in_I}, $c(N,I)=\sum_{n=N+1}^{\infty}S_{(2n)}I^{n}$
is the systematic contamination error, and $\eta$ is the random experimental
error. Our goal is to choose the intensities to minimize $|\hat{S}_{(2n)}-S_{(2n)}|$
for some desired set of orders $n$.

We start by considering the random error with $c(N,I)=0$, and consider
the noise $\eta_{k}$ on each experimental signal $S(I_{k})$ to be
independently chosen from a normal distribution with zero mean and standard deviation
$\sigma$. This choice corresponds to the common case where noise
is independent of pump intensity. We describe in Supporting Information
Section S3  how we determine $\sigma$. We use standard error propagation
techniques to determine the resulting root-mean-square random error
$r_{(2n)}=\sqrt{(\hat{S}_{(2n)}-S_{(2n)})^{2}}$ in the estimate $\hat{S}_{(2n)}$. 
In that case, $r_{(2n)}$ is determined by the rows
of $V^{-1}$ as 
\begin{equation}
r_{(2n)}=\sigma\sqrt{\sum_{j}\left(V_{nj}^{-1}\right)^{2}}.\label{eq:randomError}
\end{equation}
 When the intensities $I_{k}$ are too close to each other, making
$V$ nearly degenerate, $r_{(2n)}$ can become large.

We now consider the effects of higher-order contamination and demonstrate
how to estimate $c(N,I)$ in 2DES from intensity-dependent TA measurements.
Since TA data are faster to collect than 2DES, we use intensity-dependent
TA spectra. These TA data allow us to estimate approximate values
of both the separated orders for TA signals, $S_{(2n)}^{\text{TA}}$,
and for 2DES signals, $S_{(2n)}^{\textrm{2DES}}$, including both the
$N$ orders we want to extract and some of the neglected higher orders.
These estimates allow us to predict the optimal set of intensities
that minimize the random and systematic error. 

We use TA signals to estimate 2DES orders by exploiting a useful connection
between the two signals. When $\tau=0$, the 2DES experiment is identical
to a TA experiment, since the two pump pulses each with intensity
$I$ coherently add, giving a single pump pulse with intensity $4I$.
Therefore, $S^{\textrm{2DES}}(\tau=0,T,t,I)=S^{\textrm{TA}}(T,t,4I)$.
The factor of 4 means that, upon Taylor expanding, $S_{(2n)}^{\text{2DES}}(\tau=0,T,\omega_{t})I^{n}=4^{n}S_{(2n)}^{\text{TA}}(T,\omega_{t})I^{n}$.
We can also break the 2DES signal into its $\nQ$ pieces as 
\begin{equation}
S^{\textrm{2DES}}(\omega_{\tau},T,\omega_{t},I)=\sum_{n=-\infty}^{\infty}S^{n\textrm{Q}}(\omega_{\tau},T,\omega_{t},I),\label{eq:sum_over_nQs}
\end{equation}
where the $S^{\nQ}$ for $n<0$ are the Fourier conjugate partners
of the $S^{\nQ}$ for $n>0$ and do not contain extra information.
The $\tau=0$ 2DES signal can be obtained from a full integration
over $\omega_{\tau}$ from $-\infty$ to $\infty$, which means integrating
across all of the $\nQ$ regions. When the $\nQ$ spectra are sufficiently
spectrally separated from each other, integrating over a window at
each $\nQ$ position in $\omega_{\tau}$ in the 2D spectrum gives
$S^{\nQ}(\tau=0)$;\cite{Luttig23} the $S^{\nQ}$ can also be separated
at $\tau=0$ using phase cycling.\cite{Tan08,Maly23} Then, since
$S_{(2n)}$ only contributes to $\rQ$ when $|r|\leq n$, we have $\sum_{r=-n}^{n}S_{(2n)}^{\rQ}(\tau=0,T,\omega_{t})=S_{(2n)}^{\text{2DES}}(\tau=0,T,\omega_{t})$.
We have shown previously that, for 2DES with identical pump pulses,
$\varepsilon_{1}=\varepsilon_{2}$, \cite{Maly23,Luttig23}
\begin{equation}
S_{(2n)}^{\rQ}(\tau=0,T,\omega_{t})={2n \choose n-r}S_{(2n)}^{n\textrm{Q}}(\tau=0,T,\omega_{t}).\label{eq:nQ_relation}
\end{equation}
Noting that $\sum_{r=-n}^{n}\binom{2n}{n-r}=4^{n}$ (see Supporting
Information Section S4), we compare to Eq.~\ref{eq:sum_over_nQs}
and conclude that 
\begin{equation}
S_{(2n)}^{\text{TA}}(T,\omega_{t})=S_{(2n)}^{n\textrm{Q}}(\tau=0,T,\omega_{t}).\label{eq:TA_nQ_equality}
\end{equation}
 Equations~\ref{eq:nQ_relation} and~\ref{eq:TA_nQ_equality} allow
us to equate $S_{(2n)}^{\text{TA}}(T,\omega_{t})$ with the average
value of $S_{(2n)}^{\rQ}(\omega_{\tau},T,\omega_{t})$ over its $\omega_{\tau}$
spectral range. This connection means that if we know the contamination
$c(N,I)$ in a TA measurement, we can determine the average systematic
contamination $c(N,I)$ for each $\nQ$ region in a 2D spectrum. We
now show how to use a model for the intensity dependence of the TA
signal to produce a model for the intensity dependence of the $\nQ$
signals at $\tau=0$ via Eqs.~\ref{eq:nQ_relation} and \ref{eq:TA_nQ_equality},
which we can then use to estimate $c(N,I)$ for $\omega_{\tau}$-integrated
2D measurements. 

\begin{figure}
\includegraphics[width=3.375in]{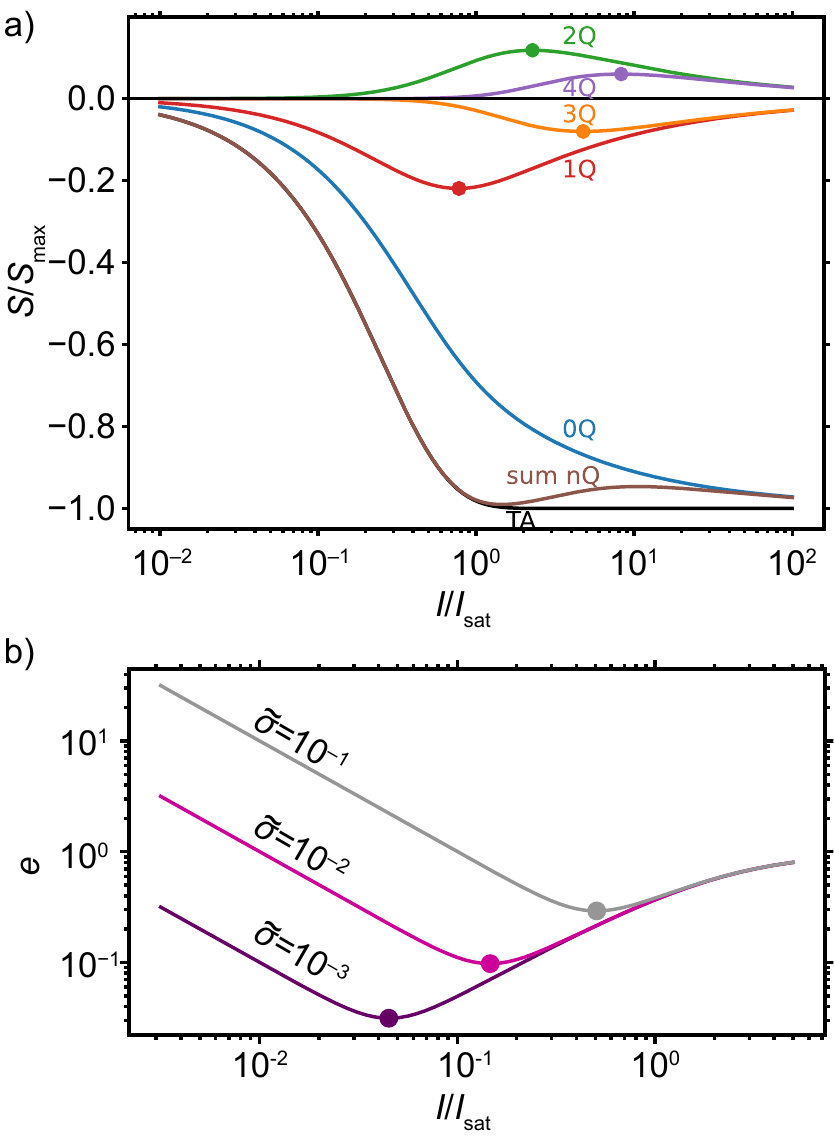}
\caption{Choice of optimal intensities. (a) Saturation behavior of TA signals
and $\protect\nQ$ signals at $\tau=0$. The TA and $0\text{Q}$ signals
saturate at the same $\protect\Smax$ while all $\protect\nQ$ signals
with $n\protect\geq1$ have maxima at finite intensity (circles). The TA signal
is plotted as $S^{\text{TA}}(4I)$ so the sum of the $\protect\nQ$
signals for all $n$ is the TA signal, as implied by Eq.~\ref{eq:sum_over_nQs}.
The sum of the $\protect\nQ$ contributions from $n=-4$ to $4$ is
shown (brown), and it agrees well with the TA curve (black)
up to $\protect\Isat$. Including larger $n$ would improve the agreement.
(b) Error in extracting lowest-order TA signal as a function of pump
intensity $I$ for three values of saturated noise-to-signal ratio
$\tilde{\sigma}=\sigma/\protect\Smax$. At low intensities, random
error dominates, while at higher intensities, the systematic error
dominates. The minimum error (circles) occurs at lower $I$ when $\tilde{\sigma}$
is smaller. \label{fig:saturation}}
\end{figure}

In many systems, TA signals saturate as pump intensity increases,
with several possible saturation forms depending on the details of
the studied system. For this discussion, we assume
\begin{equation}
S^{\textrm{TA}}(T,\omega_{t},I)=-\Smax(T,\omega_{t})\left(1-e^{-I/I_{\text{sat}}(T,\omega_{t})}\right),\label{eq:TA_sat}
\end{equation}
where $\Smax(T,\omega_{t})$ and $\Isat(T,\omega_{t})$ characterize
the exponential saturation form.\cite{diels_1996} We again suppress
the $T,~\omega_{t}$ dependence for the rest of this discussion. Using
this saturation model, we derive an analytical model (see Supporting
Information Section S5  for proof),
\begin{equation}
S^{n\textrm{Q}}(\tau=0,I)=\sum_{r=n}^{\infty}S_{(2r)}^{n\textrm{Q}}I^{r}=\Smax\begin{cases}
e^{-2I/\Isattwo}\I_{0}(2I/\Isattwo)-1 & \textrm{for }n=0,\\
(-1)^{n}e^{-2I/\Isattwo}\I_{n}(2I/\Isattwo) & \textrm{for }n\geq1,
\end{cases}\label{eq:2D_Bessel}
\end{equation}
where $\I_{n}$ is the modified Bessel function of the first kind,
with $S^{n\textrm{Q}}(\tau=0,I)$ plotted in Figure~\ref{fig:saturation}a.
This form predicts that if the TA spectrum saturates like Eq.~\ref{eq:TA_sat},
then the $\tau=0$ $\nQ$ spectrum for $n\geq1$ has a maximum amplitude
at finite $I$, unlike the TA signal itself. These maxima occur at
0.77, 2.28, 4.76, 8.26 times $\Isattwo$ for $n=1,2,3,4$, respectively.
Such maxima have not yet been observed, to our knowledge.

This exponential saturation form is frequently a good description
of TA spectra, but the method described here can be applied to any
model with a Taylor series that has a finite radius of convergence.
For example, some systems obey a saturation model called ``saturable
absorption'' in which $S^{\textrm{TA}}(I)=-\Smax\frac{I}{I+\Isat}$.\cite{Boyd08,Kumar14}
The Taylor series for saturable absorption only converges for $I<\Isat$.
In the Supporting Information Section S6.2, we show that order extraction
can also work using a saturable absorption model.

The saturation form of Eqs.~\ref{eq:TA_sat}~and~\ref{eq:2D_Bessel}
allows us to determine the measured $\hat{S}_{(n)}$ that would be
extracted from an experiment using $N$ intensities $I_{k}$, where
here we ignore the random error. Since we know the exact answers $S_{(n)}$
from the Taylor series of Eq.~\ref{eq:2D_Bessel}, we can determine
the effect of contamination error on $\hat{S}_{(n)}$. For example,
consider the $\nQ$ spectrum from Eq.~\ref{eq:2D_Bessel}. We evaluate
$S^{n\textrm{Q}}(\tau=0,I_{k})$ and use Eq.~\ref{eq:VdM_inverted}
to calculate the contaminated response orders $\hat{S}_{(2m)}$ for
$m=1,2,...,N$ from those $N$ values. The systematic error in order
$2n$ is $c_{(2n)}\equiv\hat{S}_{(2n)}-S_{(2n)}$.

We now combine the random and systematic errors by assuming that $r_{(2n)}$
and $c_{(2n)}$ are independent and add in quadrature. For either TA
spectra or $\tau=0$ $\nQ$ spectra, we define a total relative error
$e$ by taking the mean square error of $M\leq N$ orders, 
\begin{equation}
	e^{2}=\frac{1}{M}\sum_{n=1}^{M}\frac{r_{(2n)}^{2}+c_{(2n)}^{2}}{S_{(2n)}^{2}}.\label{eq:total_err}
\end{equation}
For $n$ where $S_{(2n)}$ is 0 we replace the denominator with 1,
which gives the absolute error for that term; this condition occurs,
for example, for $S^{2\textrm{Q}}(I)$ where $S_{(2)}^{\text{2Q }}=0$.
Such absolute errors depend on the choice of $I_{0}$ that defines
the intensity scale. With this definition for $e$, we then find the
set of $N$ intensities $I_{k}$ that minimize $e$. This procedure
depends on $\Isat(T,\omega_{t})$ and $\Smax(T,\omega_{t})$, so $e$
can be minimized at a particular value of $T$ and $\omega_{t}$,
or averaged over them. 

We first demonstrate this method to find the optimal intensity $I$
for extracting the lowest-order TA signal $S_{(2)}^{\text{TA }}$
with $N=1$ intensity. In this case $e$ has a simple form. The lowest-order
term of the Taylor expansion of Eq.~\ref{eq:TA_sat} is $S_{(2)}I=-S_{\text{max}}I/I_{\text{sat}}$.
For a single choice of intensity there is no Vandermonde inversion,
and $\hat{S}_{(2)}=-S_{\text{max}}(1-e^{-I/I_{\text{sat}}})/I$. Then
we find 
\begin{equation}
	e(\tilde{I})=\sqrt{\left(\frac{\tilde{\sigma}}{\tilde{I}}\right)^{2}+\left(\frac{\tilde{I}-1+\exp(-\tilde{I})}{\tilde{I}}\right)^{2}},\label{eq:totalError}
\end{equation}
where $\tilde{\sigma}\equiv\sigma/\Smax$ and $\tilde{I}\equiv I/\Isat$.
Figure~\ref{fig:saturation}b shows Eq.~\ref{eq:totalError}, along
with the optimal $\tilde{I}$, for three values of $\tilde{\sigma}$.
The optimum (minimum) occurs at higher intensity when the noise is
larger. We show the optimization of order extraction from TA spectra
with multiple pump intensities in Supporting Information Section
S6.3. We demonstrate the use of Eq.~\ref{eq:total_err} for 2D spectra
with the experimental results below.

\begin{figure}
\includegraphics[width=3.375in]{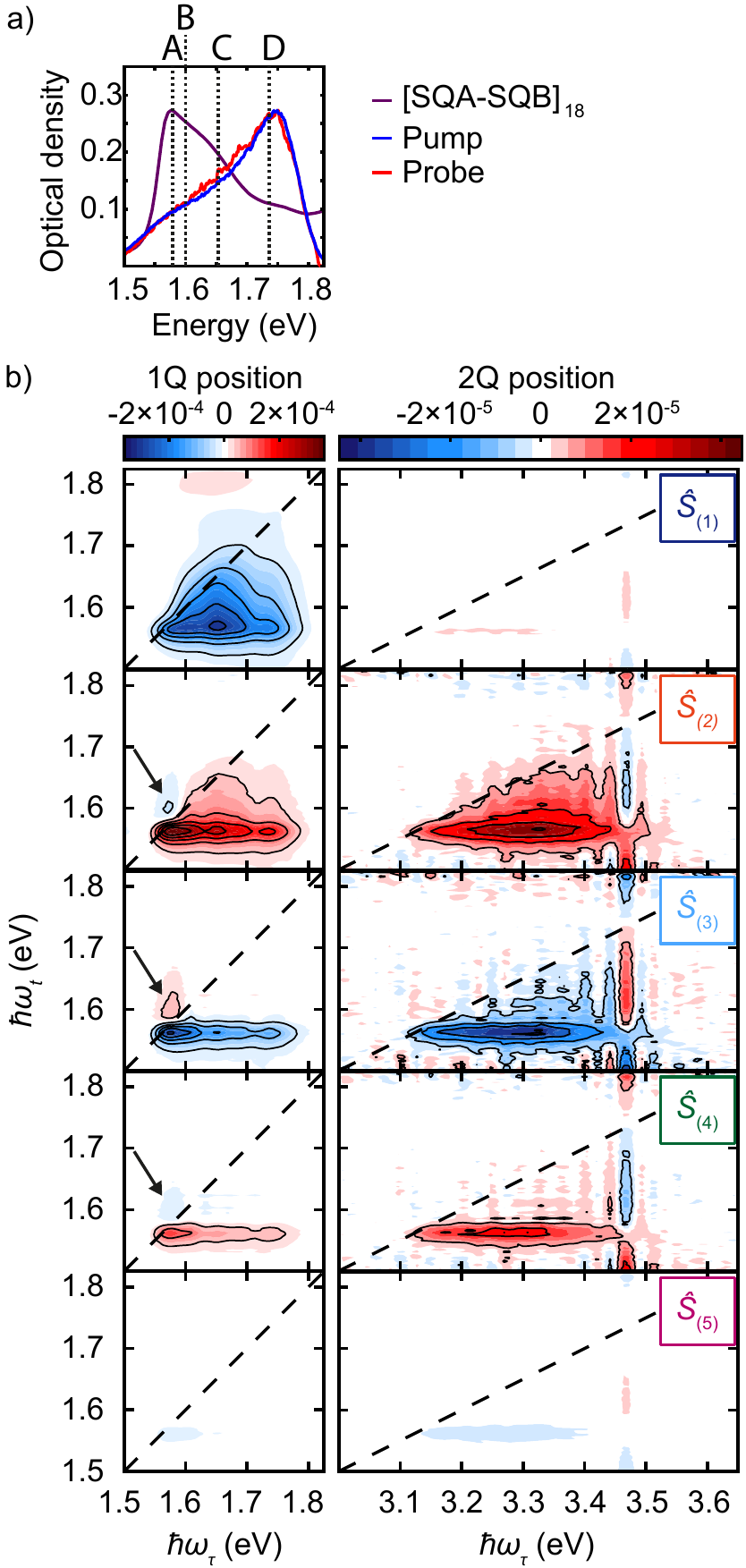}
\caption{Extracted 2D spectra. (a) Absorption spectrum of {[}SQA-SQB{]}\protect\textsubscript{18\protect\textsubscript{}}
in toluene. The vertical dashed lines indicate spectral positions
of features in the 2D spectum. (b) Extracted 1Q (left) and 2Q (right)
2D spectra of {[}SQA-SQB{]}\protect\textsubscript{18\protect\textsubscript{}}
in toluene up to $\hat{S}_{(10)}$ in the weak-probe limit. Diagonals
(black dashed lines) are drawn at $\hbar\omega_{\tau}=\hbar\omega_{t}$
for the 1Q signal (left) and $\hbar\omega_{\tau}=2\hbar\omega_{t}$
for the 2Q signal (right). Black arrows mark the NESA signals in the
higher-order 1Q spectra, as defined in the text. \label{fig:extractedOrders}}
\end{figure}

As an experimental demonstration, we perform order separation in coherently
detected 2DES of a squaraine copolymer, {[}SQA-SQB{]}\textsubscript{18\textsubscript{}},
in the weak-probe limit. The squaraine copolymer is dissolved in toluene,
leading to the formation of a J-type polymer (see absorption spectrum
in Figure~\ref{fig:extractedOrders}a).\cite{Voelker14a} To choose
our pulse intensities for 2DES, we begin with TA spectra with 101
different pulse intensities at a fixed pump-probe delay (i.e., population
time) of $T=2~\textrm{ps}$. We do not vary the pulse envelope shape
or the spot size, so we refer to pulse intensities in this work by
stating the energy per pulse. Pulse parameters and how they are determined
are discussed in Supporting Information Section S7. We fit the
dependence of the TA spectra on the pump intensity to the saturation
form of Eq.~\ref{eq:TA_sat}, finding excellent agreement (Figure~S2),
with $I_{\text{sat}}$ varying from 27~nJ to 106~nJ for $\omega_{t}$
from 1.50 to $1.76~\textrm{eV}$. We optimize the error using $\Isat$
at $\hbar\omega_{t}=A=1.58~\textrm{eV}$ corresponding to the maximum
of the absorption spectrum, which yields $I_{\text{sat}}=46~\textrm{nJ}$.
We use $\tilde{\sigma}=9\times10^{-6}$, with $\sigma=5\times10^{-7}$
and $\Smax=0.054$ as determined in Supporting Information Sections
S3 and S6.1, respectively. Table~\ref{tab:optimal_intensities}
shows the minimal error $e$ and optimal choices of intensities $I_{k}$
for 1Q and 2Q spectra where we are only interested in the accuracy
of $M=3$ orders, even when extracting $N>M$ orders. In general,
increasing $N$ improves the results. If, however, any of the optimal
intensities exceed the maximum pulse intensity experimentally available
$\Imax$, then we find an optimal $N$ to minimize the error in the
extraction, and further increases in $N$ degrade the extractions.
In our case, $I_{\text{max}}=15~\textrm{nJ}$. Our value of $I_{\text{max}}$
does not impact the error estimates for $N\leq4$, but does change
the error estimates for $N>4$, with $N=5$ being the optimal choice
for 1Q signals. We use the $N=5$ intensities from Table~\ref{tab:optimal_intensities}
that minimize the error in extracted 1Q orders, with intensities 1.2,
5.5, 10, 14, and 15~nJ. Note that while the optimal intensities are
different for the 1Q and 2Q spectra, after optimal intensities exceed
$\Imax$, those differences become negligible, giving a clear set
of intensities to use for both 1Q and 2Q spectral positions. Immediately
after the intensity-dependent TA measurements to extract $\Smax$
and $\Isat$ and random noise analysis, we carry out intensity-dependent
2D measurements while maintaining the same experimental conditions,
such as pulse overlap, spot size, and laser spectra. The intensity-dependent
2DES at 1Q and 2Q positions are shown in Figure~S4. We extract nonlinear
orders from $\hat{S}_{(2)}$ to $\hat{S}_{(10)}$ using Eq.~\ref{eq:VdM_inverted},
with results at the 1Q and 2Q positions shown in Figure~\ref{fig:extractedOrders}b,
where the dimensionless intensities $\{I_{k}\}$ are obtained by dividing
the pulse intensities by $I_{0}=\frac{1}{N}\sum_{k=1}^{N}I_{k}=9~\textrm{nJ}$ 
(see Supporting Information Section S2).
As required theoretically, the $\hat{S}_{(2)}$ contribution to the
2Q signal is small -- essentially noise (upper right panel). We compare
the extracted $\hat{S}_{(2)}$ to a low-intensity reference $\hat{S}_{L}$
in Figure~S5, which shows differences only on the noise level. 

\addtolength{\tabcolsep}{-0.2em}
\begin{table}
\begin{tabular}{clclcl}
\hline 
$\boldsymbol{N}$ &  & \textbf{1Q error} & \textbf{Optimal $I$/$\Isat$ for 1Q} & \textbf{2Q error} & \textbf{Optimal $I/\Isat$ for 2Q}\tabularnewline
\cline{1-5}
\rowcolor{gray!25}3 &  & $0.06\phantom{{0}}$ & $0.0149,0.060,0.095$ & $0.14\phantom{0}$ & 0.018, 0.073, 0.12\tabularnewline
4 &  & $0.037$ & $0.023,0.10,0.18,0.24$ & \textbf{$\mathbf{0.064}$} & 0.027, 0.12, 0.21, 0.27\tabularnewline
\rowcolor{gray!25} & $\Imax=15~\textrm{nJ}$~ & \textbf{$\mathbf{0.029}$} & 0.027, 0.12, 0.22, 0.30, 0.33 & $0.072$ & 0.027, 0.12, 0.22, 0.30, 0.33\tabularnewline
\rowcolor{gray!25}\multirow{-2}{*}{5} & $\Imax=\infty$ & $0.021$ & 0.030, 0.13, 0.26, 0.37, 0.44 & $0.040$ & 0.033, 0.15, 0.29, 0.42, 0.49\tabularnewline
\multirow{2}{*}{6} & $\Imax=15~\textrm{nJ}$~ & $0.083$ & 0.020, 0.088, 0.17, 0.25, 0.31, 0.33 & $0.21\phantom{0}$ & 0.020, 0.088, 0.17, 0.25, 0.31, 0.33\tabularnewline
 & $\Imax=\infty$ & $0.014$ & 0.035, 0.16, 0.33, 0.50, 0.63, 0.71 & $0.029$ & 0.038, 0.18, 0.35, 0.54, 0.69, 0.77\tabularnewline
\hline 
\end{tabular}\caption{Errors in extracting 1Q and 2Q spectra at $\tau=0$ using Eq.~\ref{eq:total_err}
for several choices of number of intensities~$N$. We use $\tilde{\sigma}=9\times10^{-6}$,
found in the spectra leading to Figure~\ref{fig:extractedOrders}.
We show optimal intensities for the case of unbounded $I$ and also
for the case where $I$ is not allowed to exceed $\protect\Imax=15~\textrm{nJ}$.
For $N=3$ or $N=4$, the optimal $I_{k}$ do not exceed the bound.\label{tab:optimal_intensities}
Bold values show the minimal error achievable when $I$ is bounded.}
\end{table}

The higher-order signals (Figure \ref{fig:extractedOrders}b) contain
expected features as well as new information not present in low-order
spectroscopies. With reference to the labels in Figure~\ref{fig:extractedOrders}a,
the dominant features in $\hat{S}_{(2)}^{1\text{Q}}$ are centered
at $\hbar\omega_{\tau}=C=1.65~\textrm{eV}$, slightly blueshifted
compared to the main absorption peak at $A=1.58~\textrm{eV}$, and
the dominant features in $\hat{S}_{(4)}^{2\text{Q}}$ are at $\hbar\omega_{\tau}=3.32~\textrm{eV}$,
even more blueshifted compared to twice the energy of the absorption
peak. These shifts could be due to the pump spectra, which are blue-weighted
(Figure \ref{fig:extractedOrders}a). Unlike in the raw data (Figure~S4),
at each order the signs of the dominant features are identical in
both 1Q and 2Q spectra (Figure \ref{fig:extractedOrders}), and those
signs alternate with increasing order. This alternation was previously
observed.\cite{Maly23} Reference \citenum{Rose23} explained this
sign change by showing that there are pathways contributing to $S_{2(n+1)}$
with the same $(T,\omega_{t})$ evolution as for $S_{(2n)}$ but with
opposite signs; it called these pathways ``negations,'' such as
a negated excited-state absorption (NESA) in $S_{(4)}$ that negates
the standard ESA pathway in $S_{(2)}$. The 2DES were taken at $T=2~\textrm{ps}$
(Figure~\ref{fig:extractedOrders}b), where the fifth-order 2Q signal
reaches its maximum\cite{Maly23} and energy transfer within the singly
excited manifold to the lowest-energy state has occurred.\cite{Lambert15}
The extracted 1Q spectra exhibit three prominent peaks on the excitation
axis at $\hbar\omega_{\tau}=A$, $C$, and $D$. The spectral positions
of the absolute signal maxima of $S_{(2n)}^{1\text{Q}}$ on the detection
axis lie at $\hbar\omega_{t}=A$ and show no observable shift in $\omega_{t}$
with order. However, for $\hbar\omega_{\tau}>A$, the signal with
$\hbar\omega_{t}>A$ vanishes for increasing order, leading to a narrowed
lineshape in $\omega_{t}$ direction. In contrast, the spectral position
of the absolute signal maxima of the raw data (Figure~S4) show an
apparent blueshift in $\omega_{t}$ {[}from $\hbar\omega_{t}=1.582~\textrm{eV}$
(15~nJ) to 1.569~eV (1.2~nJ){]} at high intensity, also observed
in other systems.\cite{Novoderezhkin24} This blueshift arises from
the overlap of $\hat{S}_{(2n)}$ exhibiting different signs and lineshapes
and not due to a $\omega_{t}$ shift in the $\hat{S}_{(2n)}$ themselves.
The blueshift emphasizes the importance of higher-order separation,
as even small contaminations can significantly affect the lineshape
when the lowest-order signal is desired.

The high-order 1Q response reveals a pathway hidden in the leading-order
1Q spectrum. The higher-order 1Q signals exhibit a pronounced feature
at $\hbar\omega_{\tau}=A$ and $\hbar\omega_{t}=\ensuremath{B}$ (black
arrows, Figure \ref{fig:extractedOrders}b), which is not present
in $\hat{S}{}_{(2)}^{\text{1Q }}$. In $\hat{S}{}_{(n>2)}^{\text{1Q }}$,
these features exhibit an opposite sign compared to each order's most
prominent feature, at ($A$,$A$), and they are not prominent in the
raw data at any intensity (Figure~S4). We previously demonstrated
that a negative feature appearing in $\hat{S}_{(4)}$, which is not
present in $\hat{S}_{(2)}$, can occur if 1-excitation signals are
masked in $\hat{S}_{(2)}$ due to destructive interference of signal
contributions.\cite{Rose23} Physically, one would expect an ESA peak
at ($A,B$) in the $\hat{S}_{(2)}$ signal corresponding to pump excitation
to the A exciton (Figure \ref{fig:extractedOrders}a) and probe excitation
to a biexciton state at the frequency $A+B$. Supporting Information
Section S10 uses a toy model to show that the NESA peak in $S_{(4)}^{1\textrm{Q}}$
is generally more visible than the ESA peak in $S_{(2)}^{1\textrm{Q}}$
when the A and B states arise from a weak coupling. While this manuscript
is focused on the introduction of the high-order 2DES technique itself,
this discussion shows just one piece of useful spectral information
visible in the high-order and not in the leading-order spectra.

\begin{figure}
\includegraphics[width=3.375in]{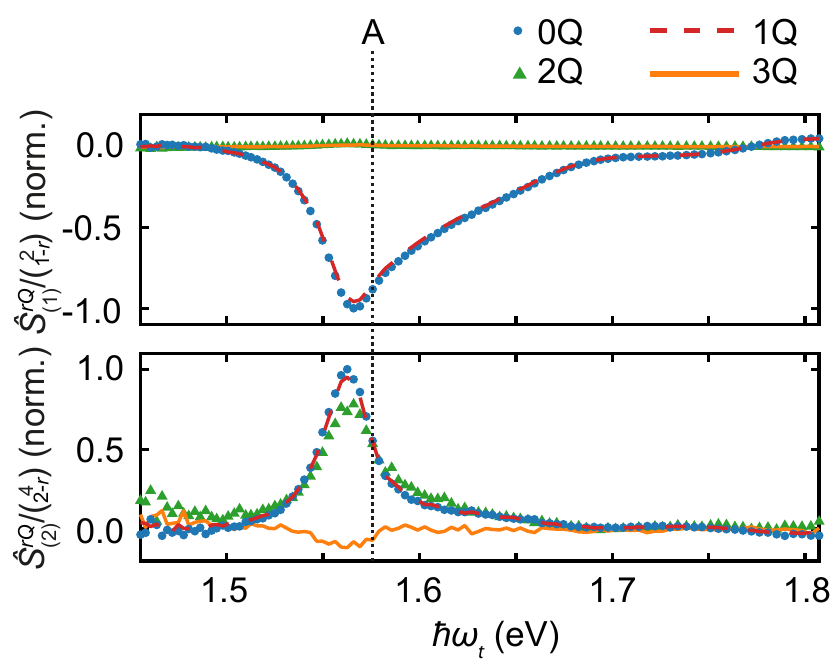}
\caption{Self-consistency checks. $\hat{S}_{(2n)}^{\protect\rQ}/{2n \choose n-r}$
after window-integrating over $\omega_{\tau}$. When $r>n$, $\hat{S}_{(2n)}^{\protect\rQ}$
signals should be zero and are not scaled by ${2n \choose n-r}$.
The integration windows are defined as $\text{\ensuremath{\omega_{0}}}[r-1/2,r+1/2]$
for $\omega_{0}=1.59$~eV. For each $n=1,2$, signals are normalized
with respect to the absolute maximum values of $\hat{S}_{(2n)}^{0\textrm{Q}}$.
The absorption maximum, marked by A as in Fig.~\ref{fig:extractedOrders}a, 
is the frequency at which $\Smax$ and $\Isat$ were extracted when optimizing
the pump intensities and is the frequency at which the errors in 
Table~\ref{tab:optimal_intensities} were estimated.
\label{fig:selfConsistency}}
\end{figure}

We now outline a procedure to verify that the response orders $\hat{S}_{(n)}$
have been correctly extracted with the chosen set of intensities $\{I_{k}\}$,
independent of the model-based error analysis presented above. Equation~\ref{eq:nQ_relation}
shows that the extracted orders $\hat{S}_{(2n)}^{r\textrm{Q}}(\tau=0,T,\omega_{t})$/${2n \choose n-r}$
with fixed order $n$, should be equal at all $T$ and $\omega_{t}$,
if $r\leq n$. Signals with $r>n$ should be zero. If these relations
do not hold, it is model-independent evidence of error in $\hat{S}_{(2n)}$.
We find $\hat{S}_{(2n)}^{r\textrm{Q}}(\tau=0)$ by integrating $\hat{S}_{(2n)}^{r\textrm{Q}}(\omega_{\tau})$,
divide by ${2n \choose n-r}$ if $r\leq n$, and plot the results
in Figure~\ref{fig:selfConsistency}. Fulfilling the self-consistency
check, the $\hat{S}_{(2)}^{\rQ}$ for $r\geq2$ are zero (Figure \ref{fig:selfConsistency}
top, green and orange curves), while the normalized 0Q and 1Q signals
are equal (Figure \ref{fig:selfConsistency} top, blue and red curves).
Similar to the lowest-order signal, the integrated and normalized
extracted orders $\hat{S}_{(4)}^{0\textrm{Q}}$ and $\hat{S}_{(4)}^{1\textrm{Q}}$
overlap nearly perfectly (Figure \ref{fig:selfConsistency} bottom,
blue and red curves). The $\hat{S}_{(4)}^{2\textrm{Q}}$ signal shows
some systematic deviation from the 0Q and 1Q (Figure \ref{fig:selfConsistency}
bottom, green curve), with a slight decrease of the peak at 1.56~eV
and an increased blue shoulder indicating a small systematic error.
One potential source of the remaining error comes from fluctuations
of the applied pulse intensities. The $\hat{S}_{(4)}^{3\textrm{Q}}$
signal (Figure \ref{fig:selfConsistency} bottom, orange curve) is
supposed to be zero but shows a small deviation near 1.57~eV, indicative
of contamination. Overall, the self-consistency checks demonstrate
reasonable extraction of the nonlinear signals and support the conclusion
that the lowest-order extractions $\hat{S}_{(2)}^{\rQ}$\textbf{ }are
essentially uncontaminated, while $\hat{S}_{(4)}^{\rQ}$\textbf{ }have
some contamination.

All signals in nonlinear spectroscopy strongly rely on the intensities
of the excitation pulses. While a sufficiently high pulse intensity
is necessary to measure a signal stronger than the noise level, higher-order
terms that may be undesired can contribute to the measured signal.
Here, we generalized the recently developed intensity-cycling method
to not require specific intensity ratios between the spectra. As a
result, the presented order-separation technique can be applied to
previously collected fluence-dependent spectra. In addition, the new
scheme is applicable not only to TA data, as in the previously reported
extraction procedure, but can also be used to separate nonlinear orders
in 2DES. We showed how to use the connection between TA spectra and
excitation-frequency-integrated 2DES spectra to estimate the systematic
errors in a high-order extraction and thus to choose optimal pump
intensities for the extraction. We demonstrated the method on coherently
detected 2DES measurements on squaraine copolymers {[}SQA-SQB{]}\textsubscript{18\textsubscript{}}
in the weak-probe limit and showed that the high-order 1Q spectra
reveal an ESA peak that is masked in the lowest-order spectrum. While
we focus in this publication on TA and coherently detected 2D spectroscopy,
the procedure presented here is readily adaptable to other spectroscopies
such as transient grating, action-detected 2DES, or time-resolved
photoemission spectroscopy.\cite{Kahl18,dabrowski_ultrafast_2020,Huber19,lv_angle-resolved_2019,Vogelsang24}

\section{Experimental Methods}

The intensities and delays of the pump pulses were altered by an acousto-optic
modulator pulse shaper on a shot-to-shot basis, which also chopped
every second pump pulse, in a partially non-collinear pump-probe setup.
The pump beam was blocked after the sample, and the probe beam was
detected shot-to-shot via a spectrometer and line camera. For the
correction of the 2D data, five 2D measurements with pump pulse energies
(1.2, 5.5, 10, 14, 15)~nJ were taken with the pulse shaper maintaining
the same pulse envelope for each $\tau$ step. Simultaneously we measured
a low-excitation-intensity reference at 0.16~nJ. All 2D spectra were
taken with 299 steps in $\tau$ of step size 0.37~fs at a population
time of $T=2~\textrm{ps}$. For every average, all six 2D measurements
were taken consecutively before repeating the pulse sequence. For
setup details and signal processing see Supporting Information
Sections S7 and S8.

\section{Data Availability}
Data for all figures are available on Zenodo.\cite{Krich25a}

\section{Supporting Information}
Supporting Information: Molecular structure, effects of choosing the base power, random error, connection between TA and 2DES signals, analytical saturation model for 2DES signals, choosing optimal intensities, experimental setup, raw data and data processing, comparison of lowest order with low-intensity reference measurement, and visibility enhancement of cross peaks in higher-order signals.

\begin{acknowledgement}
We acknowledge funding by the European Research Council (ERC) within
Advanced Grant IMPACTS (No. 101141366) and from the Natural Sciences
and Engineering Research Council of Canada (NSERC), {[}597081-24{]}. 
P.~M.\ acknowledges funding by Charles University (grant PRIMUS/24/SCI/007).
We further thank Christoph Lambert and Arthur Turkin for providing
the sample and its synthesis and Federico Gallina and Hugo Lafleur for careful
reading of the manuscript.
\end{acknowledgement}
\bibliography{NLO}

\providecommand{\latin}[1]{#1}
\makeatletter
\providecommand{\doi}
  {\begingroup\let\do\@makeother\dospecials
  \catcode`\{=1 \catcode`\}=2 \doi@aux}
\providecommand{\doi@aux}[1]{\endgroup\texttt{#1}}
\makeatother
\providecommand*\mcitethebibliography{\thebibliography}
\csname @ifundefined\endcsname{endmcitethebibliography}
  {\let\endmcitethebibliography\endthebibliography}{}
\begin{mcitethebibliography}{33}
\providecommand*\natexlab[1]{#1}
\providecommand*\mciteSetBstSublistMode[1]{}
\providecommand*\mciteSetBstMaxWidthForm[2]{}
\providecommand*\mciteBstWouldAddEndPuncttrue
  {\def\EndOfBibitem{\unskip.}}
\providecommand*\mciteBstWouldAddEndPunctfalse
  {\let\EndOfBibitem\relax}
\providecommand*\mciteSetBstMidEndSepPunct[3]{}
\providecommand*\mciteSetBstSublistLabelBeginEnd[3]{}
\providecommand*\EndOfBibitem{}
\mciteSetBstSublistMode{f}
\mciteSetBstMaxWidthForm{subitem}{(\alph{mcitesubitemcount})}
\mciteSetBstSublistLabelBeginEnd
  {\mcitemaxwidthsubitemform\space}
  {\relax}
  {\relax}

\bibitem[Mukamel(1999)]{MukamelBook}
Mukamel,~S. \emph{Principles of nonlinear optical spectroscopy}; Oxford,
  1999\relax
\mciteBstWouldAddEndPuncttrue
\mciteSetBstMidEndSepPunct{\mcitedefaultmidpunct}
{\mcitedefaultendpunct}{\mcitedefaultseppunct}\relax
\EndOfBibitem
\bibitem[Mal{\'{y}} \latin{et~al.}(2023)Mal{\'{y}}, L\"uttig, Rose, Turkin,
  Lambert, Krich, and Brixner]{Maly23}
Mal{\'{y}},~P.; L\"uttig,~J.; Rose,~P.~A.; Turkin,~A.; Lambert,~C.;
  Krich,~J.~J.; Brixner,~T. Separating single- from multi-particle dynamics in
  nonlinear spectroscopy. \emph{Nature} \textbf{2023}, \emph{616},
  280--287\relax
\mciteBstWouldAddEndPuncttrue
\mciteSetBstMidEndSepPunct{\mcitedefaultmidpunct}
{\mcitedefaultendpunct}{\mcitedefaultseppunct}\relax
\EndOfBibitem
\bibitem[L{\"u}ttig \latin{et~al.}(2023)L{\"u}ttig, Rose, Mal{\'{y}}, Turkin,
  B{\"u}hler, Lambert, Krich, and Brixner]{Luttig23}
L{\"u}ttig,~J.; Rose,~P.~A.; Mal{\'{y}},~P.; Turkin,~A.; B{\"u}hler,~M.;
  Lambert,~C.; Krich,~J.~J.; Brixner,~T. High-order pump-probe and high-order
  two-dimensional electronic spectroscopy on the example of squaraine
  oligomers. \emph{J. Chem. Phys.} \textbf{2023}, \emph{158}, 234201\relax
\mciteBstWouldAddEndPuncttrue
\mciteSetBstMidEndSepPunct{\mcitedefaultmidpunct}
{\mcitedefaultendpunct}{\mcitedefaultseppunct}\relax
\EndOfBibitem
\bibitem[Rose and Krich(2023)Rose, and Krich]{Rose23}
Rose,~P.~A.; Krich,~J.~J. Interpretations of high-order transient absorption
  spectroscopies. \emph{J. Phys. Chem. Lett.} \textbf{2023}, \emph{14},
  10849--10855\relax
\mciteBstWouldAddEndPuncttrue
\mciteSetBstMidEndSepPunct{\mcitedefaultmidpunct}
{\mcitedefaultendpunct}{\mcitedefaultseppunct}\relax
\EndOfBibitem
\bibitem[Heshmatpour \latin{et~al.}(2020)Heshmatpour, Malevich, Plasser,
  Menger, Lambert, \v{S}anda, and Hauer]{Heshmatpour20}
Heshmatpour,~C.; Malevich,~P.; Plasser,~F.; Menger,~M.; Lambert,~C.;
  \v{S}anda,~F.; Hauer,~J. Annihilation dynamics of molecular excitons measured
  at a single perturbative excitation energy. \emph{J. Phys. Chem. Lett.}
  \textbf{2020}, \emph{11}, 7776--7781\relax
\mciteBstWouldAddEndPuncttrue
\mciteSetBstMidEndSepPunct{\mcitedefaultmidpunct}
{\mcitedefaultendpunct}{\mcitedefaultseppunct}\relax
\EndOfBibitem
\bibitem[L\"uttig \latin{et~al.}(2023)L\"uttig, Mueller, Mal{\'{y}}, Krich, and
  Brixner]{Luttig23a}
L\"uttig,~J.; Mueller,~S.; Mal{\'{y}},~P.; Krich,~J.~J.; Brixner,~T.
  Higher-order multidimensional and pump{\textendash}probe spectroscopies.
  \emph{J. Phys. Chem. Lett.} \textbf{2023}, \emph{14}, 7556--7573\relax
\mciteBstWouldAddEndPuncttrue
\mciteSetBstMidEndSepPunct{\mcitedefaultmidpunct}
{\mcitedefaultendpunct}{\mcitedefaultseppunct}\relax
\EndOfBibitem
\bibitem[Ding \latin{et~al.}(2005)Ding, Fulmer, and
  Zanni]{ding_heterodyned_2005}
Ding,~F.; Fulmer,~E.~C.; Zanni,~M.~T. Heterodyned fifth-order two-dimensional
  {IR} spectroscopy: {Third}-quantum states and polarization selectivity.
  \emph{J. Chem. Phys.} \textbf{2005}, \emph{123}, 094502\relax
\mciteBstWouldAddEndPuncttrue
\mciteSetBstMidEndSepPunct{\mcitedefaultmidpunct}
{\mcitedefaultendpunct}{\mcitedefaultseppunct}\relax
\EndOfBibitem
\bibitem[Yu \latin{et~al.}(2018)Yu, Titze, Zhu, Liu, and
  Li]{yu_observation_2018}
Yu,~S.; Titze,~M.; Zhu,~Y.; Liu,~X.; Li,~H. Observation of scalable and
  deterministic multi-atom {Dicke} states in an atomic vapor.
  \emph{arXiv:1807.09300 [physics]} \textbf{2018}, arXiv: 1807.09300\relax
\mciteBstWouldAddEndPuncttrue
\mciteSetBstMidEndSepPunct{\mcitedefaultmidpunct}
{\mcitedefaultendpunct}{\mcitedefaultseppunct}\relax
\EndOfBibitem
\bibitem[Mal{\'{y}} \latin{et~al.}(2020)Mal{\'{y}}, L\"uttig, Turkin,
  Dost{\'{a}}l, Lambert, and Brixner]{Maly20}
Mal{\'{y}},~P.; L\"uttig,~J.; Turkin,~A.; Dost{\'{a}}l,~J.; Lambert,~C.;
  Brixner,~T. From wavelike to sub-diffusive motion: exciton dynamics and
  interaction in squaraine copolymers of varying length. \emph{Chem. Sci.}
  \textbf{2020}, \emph{11}, 456--466\relax
\mciteBstWouldAddEndPuncttrue
\mciteSetBstMidEndSepPunct{\mcitedefaultmidpunct}
{\mcitedefaultendpunct}{\mcitedefaultseppunct}\relax
\EndOfBibitem
\bibitem[Br\"uggemann and Pullerits(2011)Br\"uggemann, and
  Pullerits]{bruggemann_nonperturbative_2011}
Br\"uggemann,~B.; Pullerits,~T. Nonperturbative modeling of fifth-order
  coherent multidimensional spectroscopy in light harvesting antennas.
  \emph{New J. Phys.} \textbf{2011}, \emph{13}, 025024\relax
\mciteBstWouldAddEndPuncttrue
\mciteSetBstMidEndSepPunct{\mcitedefaultmidpunct}
{\mcitedefaultendpunct}{\mcitedefaultseppunct}\relax
\EndOfBibitem
\bibitem[Turner and Nelson(2010)Turner, and Nelson]{Turner10}
Turner,~D.~B.; Nelson,~K.~A. Coherent measurements of high-order electronic
  correlations in quantum wells. \emph{Nature} \textbf{2010}, \emph{466},
  1089--1092\relax
\mciteBstWouldAddEndPuncttrue
\mciteSetBstMidEndSepPunct{\mcitedefaultmidpunct}
{\mcitedefaultendpunct}{\mcitedefaultseppunct}\relax
\EndOfBibitem
\bibitem[Brixner \latin{et~al.}(2005)Brixner, Stenger, Vaswani, Cho,
  Blankenship, and Fleming]{Brixner05}
Brixner,~T.; Stenger,~J.; Vaswani,~H.~M.; Cho,~M.; Blankenship,~R.~E.;
  Fleming,~G.~R. Two-dimensional spectroscopy of electronic couplings in
  photosynthesis. \emph{Nature} \textbf{2005}, \emph{434}, 625--628\relax
\mciteBstWouldAddEndPuncttrue
\mciteSetBstMidEndSepPunct{\mcitedefaultmidpunct}
{\mcitedefaultendpunct}{\mcitedefaultseppunct}\relax
\EndOfBibitem
\bibitem[Oliver \latin{et~al.}(2014)Oliver, Lewis, and
  Fleming]{oliver_correlating_2014}
Oliver,~T. A.~A.; Lewis,~N. H.~C.; Fleming,~G.~R. Correlating the motion of
  electrons and nuclei with two-dimensional electronic{\textendash}vibrational
  spectroscopy. \emph{PNAS} \textbf{2014}, \emph{111}, 10061--10066\relax
\mciteBstWouldAddEndPuncttrue
\mciteSetBstMidEndSepPunct{\mcitedefaultmidpunct}
{\mcitedefaultendpunct}{\mcitedefaultseppunct}\relax
\EndOfBibitem
\bibitem[Turner \latin{et~al.}(2011)Turner, Wilk, Curmi, and Scholes]{Turner11}
Turner,~D.~B.; Wilk,~K.~E.; Curmi,~P. M.~G.; Scholes,~G.~D. Comparison of
  Electronic and Vibrational Coherence Measured by Two-Dimensional Electronic
  Spectroscopy. \emph{J. Phys. Chem. Lett.} \textbf{2011}, \emph{2},
  1904--1911\relax
\mciteBstWouldAddEndPuncttrue
\mciteSetBstMidEndSepPunct{\mcitedefaultmidpunct}
{\mcitedefaultendpunct}{\mcitedefaultseppunct}\relax
\EndOfBibitem
\bibitem[Fuller and Ogilvie(2015)Fuller, and Ogilvie]{Fuller15}
Fuller,~F.~D.; Ogilvie,~J.~P. Experimental implementations of two-dimensional
  {Fourier} transform electronic spectroscopy. \emph{Annu. Rev. Phys. Chem.}
  \textbf{2015}, \emph{66}, 667--690\relax
\mciteBstWouldAddEndPuncttrue
\mciteSetBstMidEndSepPunct{\mcitedefaultmidpunct}
{\mcitedefaultendpunct}{\mcitedefaultseppunct}\relax
\EndOfBibitem
\bibitem[Dost{\'{a}}l \latin{et~al.}(2018)Dost{\'{a}}l, Fennel, Koch, Herbst,
  W{\"u}rthner, and Brixner]{Dostal18}
Dost{\'{a}}l,~J.; Fennel,~F.; Koch,~F.; Herbst,~S.; W{\"u}rthner,~F.;
  Brixner,~T. Direct observation of exciton{\textendash}exciton interactions.
  \emph{Nat. Commun.} \textbf{2018}, \emph{9}, 2466\relax
\mciteBstWouldAddEndPuncttrue
\mciteSetBstMidEndSepPunct{\mcitedefaultmidpunct}
{\mcitedefaultendpunct}{\mcitedefaultseppunct}\relax
\EndOfBibitem
\bibitem[V{\"o}lker \latin{et~al.}(2014)V{\"o}lker, Dellermann, Ceymann,
  Holzapfel, and Lambert]{Voelker2014}
V{\"o}lker,~S.~F.; Dellermann,~T.; Ceymann,~H.; Holzapfel,~M.; Lambert,~C.
  Synthesis, electrochemical, and optical properties of low band gap homo- and
  copolymers based on squaraine dyes. \emph{J. Polym. Sci. Part A: Polym.
  Chem.} \textbf{2014}, \emph{52}, 890--911\relax
\mciteBstWouldAddEndPuncttrue
\mciteSetBstMidEndSepPunct{\mcitedefaultmidpunct}
{\mcitedefaultendpunct}{\mcitedefaultseppunct}\relax
\EndOfBibitem
\bibitem[Shim and Zanni(2009)Shim, and Zanni]{Shim09}
Shim,~S.-H.; Zanni,~M.~T. How to turn your pump-probe instrument into a
  multidimensional spectrometer: 2D IR and Vis spectroscopiesvia pulse shaping.
  \emph{Phys. Chem. Chem. Phys.} \textbf{2009}, \emph{11}, 748--761\relax
\mciteBstWouldAddEndPuncttrue
\mciteSetBstMidEndSepPunct{\mcitedefaultmidpunct}
{\mcitedefaultendpunct}{\mcitedefaultseppunct}\relax
\EndOfBibitem
\bibitem[Blank \latin{et~al.}(1999)Blank, Kaufman, and Fleming]{Blank99}
Blank,~D.~A.; Kaufman,~L.~J.; Fleming,~G.~R. Fifth-order two-dimensional Raman
  spectra of CS2 are dominated by third-order cascades. \emph{J. Chem. Phys.}
  \textbf{1999}, \emph{111}, 3105--3114\relax
\mciteBstWouldAddEndPuncttrue
\mciteSetBstMidEndSepPunct{\mcitedefaultmidpunct}
{\mcitedefaultendpunct}{\mcitedefaultseppunct}\relax
\EndOfBibitem
\bibitem[Tan(2008)]{Tan08}
Tan,~H.-S. Theory and phase-cycling scheme selection principles of collinear
  phase coherent multi-dimensional optical spectroscopy. \emph{J. Chem. Phys.}
  \textbf{2008}, \emph{129}, 124501\relax
\mciteBstWouldAddEndPuncttrue
\mciteSetBstMidEndSepPunct{\mcitedefaultmidpunct}
{\mcitedefaultendpunct}{\mcitedefaultseppunct}\relax
\EndOfBibitem
\bibitem[Diels and Rudolph(1996)Diels, and Rudolph]{diels_1996}
Diels,~J.-C.; Rudolph,~W. \emph{Ultrashort laser pulse phenomena: fundamentals,
  techniques, and applications on a femtosecond time scale}, 2nd ed.; Academic
  Press Inc: Amsterdam, 1996\relax
\mciteBstWouldAddEndPuncttrue
\mciteSetBstMidEndSepPunct{\mcitedefaultmidpunct}
{\mcitedefaultendpunct}{\mcitedefaultseppunct}\relax
\EndOfBibitem
\bibitem[Boyd(2008)]{Boyd08}
Boyd,~R. \emph{Nonlinear Optics}; Academic Press, 2008\relax
\mciteBstWouldAddEndPuncttrue
\mciteSetBstMidEndSepPunct{\mcitedefaultmidpunct}
{\mcitedefaultendpunct}{\mcitedefaultseppunct}\relax
\EndOfBibitem
\bibitem[Kumar \latin{et~al.}(2014)Kumar, Cui, Ceballos, He, Wang, and
  Zhao]{Kumar14}
Kumar,~N.; Cui,~Q.; Ceballos,~F.; He,~D.; Wang,~Y.; Zhao,~H. Exciton-exciton
  annihilation in {MoSe}${}_{2}$ monolayers. \emph{Phys. Rev. B} \textbf{2014},
  \emph{89}, 125427\relax
\mciteBstWouldAddEndPuncttrue
\mciteSetBstMidEndSepPunct{\mcitedefaultmidpunct}
{\mcitedefaultendpunct}{\mcitedefaultseppunct}\relax
\EndOfBibitem
\bibitem[V{\"o}lker \latin{et~al.}(2014)V{\"o}lker, Schmiedel, Holzapfel,
  Renziehausen, Engel, and Lambert]{Voelker14a}
V{\"o}lker,~S.~F.; Schmiedel,~A.; Holzapfel,~M.; Renziehausen,~K.; Engel,~V.;
  Lambert,~C. Singlet-Singlet Exciton Annihilation in an Exciton-Coupled
  Squaraine-Squaraine Copolymer: A Model toward Hetero-J-Aggregates. \emph{J.
  Phys. Chem. C} \textbf{2014}, \emph{118}, 17467--17482\relax
\mciteBstWouldAddEndPuncttrue
\mciteSetBstMidEndSepPunct{\mcitedefaultmidpunct}
{\mcitedefaultendpunct}{\mcitedefaultseppunct}\relax
\EndOfBibitem
\bibitem[Lambert \latin{et~al.}(2015)Lambert, Koch, V\"olker, Schmiedel,
  Holzapfel, Humeniuk, R\"ohr, Mitric, and Brixner]{Lambert15}
Lambert,~C.; Koch,~F.; V\"olker,~S.~F.; Schmiedel,~A.; Holzapfel,~M.;
  Humeniuk,~A.; R\"ohr,~M. I.~S.; Mitric,~R.; Brixner,~T. Energy transfer
  between squaraine polymer sections: {From} helix to zigzag and all the way
  back. \emph{J. Am. Chem. Soc.} \textbf{2015}, \emph{137}, 7851--7861\relax
\mciteBstWouldAddEndPuncttrue
\mciteSetBstMidEndSepPunct{\mcitedefaultmidpunct}
{\mcitedefaultendpunct}{\mcitedefaultseppunct}\relax
\EndOfBibitem
\bibitem[Novoderezhkin and Razjivin(2024)Novoderezhkin, and
  Razjivin]{Novoderezhkin24}
Novoderezhkin,~V.~I.; Razjivin,~A.~P. Multiexciton spectra of molecular
  aggregates: application to photosynthetic antenna complexes. \emph{Phys.
  Chem. Chem. Phys.} \textbf{2024}, \emph{26}, 23800--23810\relax
\mciteBstWouldAddEndPuncttrue
\mciteSetBstMidEndSepPunct{\mcitedefaultmidpunct}
{\mcitedefaultendpunct}{\mcitedefaultseppunct}\relax
\EndOfBibitem
\bibitem[Kahl \latin{et~al.}(2018)Kahl, Podbiel, Schneider, Makris, Sindermann,
  Witt, Kilbane, Hoegen, Aeschlimann, and zu~Heringdorf]{Kahl18}
Kahl,~P.; Podbiel,~D.; Schneider,~C.; Makris,~A.; Sindermann,~S.; Witt,~C.;
  Kilbane,~D.; Hoegen,~M. H.-v.; Aeschlimann,~M.; zu~Heringdorf,~F.~M. Direct
  observation of surface plasmon polariton propagation and interference by
  time-resolved imaging in normal-incidence two photon photoemission
  microscopy. \emph{Plasmonics} \textbf{2018}, \emph{13}, 239--246\relax
\mciteBstWouldAddEndPuncttrue
\mciteSetBstMidEndSepPunct{\mcitedefaultmidpunct}
{\mcitedefaultendpunct}{\mcitedefaultseppunct}\relax
\EndOfBibitem
\bibitem[D\c{a}browski \latin{et~al.}(2020)D\c{a}browski, Dai, and
  Petek]{dabrowski_ultrafast_2020}
D\c{a}browski,~M.; Dai,~Y.; Petek,~H. Ultrafast photoemission electron
  microscopy: {Imaging} plasmons in space and time. \emph{Chem. Rev.}
  \textbf{2020}, \emph{120}, 6247--6287\relax
\mciteBstWouldAddEndPuncttrue
\mciteSetBstMidEndSepPunct{\mcitedefaultmidpunct}
{\mcitedefaultendpunct}{\mcitedefaultseppunct}\relax
\EndOfBibitem
\bibitem[Huber \latin{et~al.}(2019)Huber, Pres, Wittmann, Dietrich, L\"uttig,
  Fersch, Krauss, Friedrich, Kern, Lisinetskii, Hensen, Hecht, Bratschitsch,
  Riedle, and Brixner]{Huber19}
Huber,~B.; Pres,~S.; Wittmann,~E.; Dietrich,~L.; L\"uttig,~J.; Fersch,~D.;
  Krauss,~E.; Friedrich,~D.; Kern,~J.; Lisinetskii,~V. \latin{et~al.}  {Space-
  and time-resolved UV-to-NIR surface spectroscopy and 2D nanoscopy at 1 MHz
  repetition rate}. \emph{Rev. Sci. Instrum.} \textbf{2019}, \emph{90},
  113103\relax
\mciteBstWouldAddEndPuncttrue
\mciteSetBstMidEndSepPunct{\mcitedefaultmidpunct}
{\mcitedefaultendpunct}{\mcitedefaultseppunct}\relax
\EndOfBibitem
\bibitem[Lv \latin{et~al.}(2019)Lv, Qian, and Ding]{lv_angle-resolved_2019}
Lv,~B.; Qian,~T.; Ding,~H. Angle-resolved photoemission spectroscopy and its
  application to topological materials. \emph{Nat. Rev. Phys.} \textbf{2019},
  \emph{1}, 609--626\relax
\mciteBstWouldAddEndPuncttrue
\mciteSetBstMidEndSepPunct{\mcitedefaultmidpunct}
{\mcitedefaultendpunct}{\mcitedefaultseppunct}\relax
\EndOfBibitem
\bibitem[Vogelsang \latin{et~al.}(2024)Vogelsang, Wittenbecher, Mikaelsson,
  Guo, Sytcevich, Viotti, Arnold, L'Huillier, and Mikkelsen]{Vogelsang24}
Vogelsang,~J.; Wittenbecher,~L.; Mikaelsson,~S.; Guo,~C.; Sytcevich,~I.;
  Viotti,~A.-L.; Arnold,~C.~L.; L'Huillier,~A.; Mikkelsen,~A. Time-resolved
  photoemission electron microscopy on a {ZnO} surface using an extreme
  ultraviolet attosecond pulse pair. \emph{Adv. Phys. Res.} \textbf{2024},
  \emph{3}, 2300122\relax
\mciteBstWouldAddEndPuncttrue
\mciteSetBstMidEndSepPunct{\mcitedefaultmidpunct}
{\mcitedefaultendpunct}{\mcitedefaultseppunct}\relax
\EndOfBibitem
\bibitem[Krich \latin{et~al.}(2025)Krich, Brenneis, and Rose]{Krich25a}
Krich,~J.~J.; Brenneis,~L.; Rose,~P.~A. Data for figures in "Separating orders
  of response in transient absorption and coherent multi-dimensional
  spectroscopy by intensity variation". Zenodo, 2025;
  \url{https://doi.org/10.5281/zenodo.15519457}\relax
\mciteBstWouldAddEndPuncttrue
\mciteSetBstMidEndSepPunct{\mcitedefaultmidpunct}
{\mcitedefaultendpunct}{\mcitedefaultseppunct}\relax
\EndOfBibitem
\end{mcitethebibliography}

\vfill\eject


\section*{Supplementary material}

# Supporting Information for Separating orders of response in transient absorption and coherent multi-dimensional spectroscopy by intensity variation

Jacob J. Krich,*,†,‡,# Luisa Brenneis,¶,# Peter A. Rose,† Katja Mayershofer,¶ Simon Büttner,¶ Julian Lüttig,§ Pavel Malý,∥ and Tobias Brixner*,¶,⊥

†*Department of Physics, University of Ottawa, Ottawa, ON K1N 6N5, Canada*
‡*Nexus for Quantum Technologies, University of Ottawa, Ottawa, ON K1N 6N5, Canada*
¶*Institut für Physikalische und Theoretische Chemie, Universität Würzburg, Am Hubland, 97074 Würzburg, Germany*
§*Department of Physics, University of Michigan, 450 Church St., Ann Arbor, MI 48109, USA*
∥*Faculty of Mathematics and Physics, Charles University, Ke Karlovu 5, 121 16 Prague, Czech Republic*
⊥*Center for Nanosystems Chemistry (CNC), Universität Würzburg, Theodor-Boveri-Weg, 97074 Würzburg, Germany*
#*Contributed equally to this work*

E-mail: jkrich@uottawa.ca; tobias.brixner@uni-wuerzburg.de

## S1 Molecular structure

The experimental sample is a squaraine copolymer [SQA-SQB]$_{18}$ (Figure S1) dissolved in toluene. On average, the copolymer consists of 18 [SQA-SQB] dimer units; for details on the degree of polymerization, see Reference 1. The synthesis of the copolymer was carried out as described previously.[1,2]

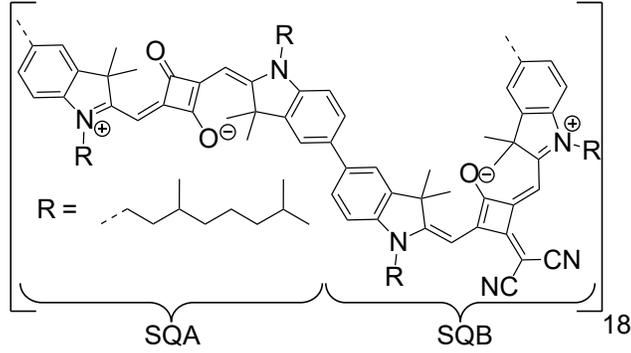

Figure S1: Chemical structure of the copolymer [SQA-SQB]$_{18}$, consisting on average of 18 [SQA-SQB] dimer units.

## S2 Effects of choosing the base power

In the main text we discuss the order extraction using unitless intensities, as in Eq. 5. In such a case, the extracted orders $\hat{S}_{(n)}$ all have the same units. Nonetheless, the extracted orders depend on the choice of what intensity is labeled to be 1. Consider that we choose to scale intensities with the dimensionful intensity $I_0$. Then a pulse with dimensionful intensity $\bar{I}$ has dimensionless intensity $I = \bar{I}/I_0$. We here briefly discuss the impact of the choice of base intensity $I_0$. Reproducing Eq. 2 from the main text here,

$$S(\tau, T, t, I) = \sum_{j=1}^{\infty} S_{(2j)}(\tau, T, t) I^j, \tag{S1}$$

$$S(\tau, T, t, \bar{I}) = \sum_{j=1}^{\infty} S_{(2j)}(\tau, T, t) \left(\frac{\bar{I}}{I_0}\right)^j. \tag{S2}$$

We define

$$s_{(2j)} I_0^j = S_{(2j)}, \tag{S3}$$

which gives

$$S(\tau, T, t, I) = \sum_{j=1}^{\infty} s_{(2j)}(\tau, T, t) \bar{I}^j, \tag{S4}$$

which yields the expansion in terms of intensities with units. This form demonstrates that $S_{(2j)} \propto I_0^j$, and therefore the choice of $I_0$ scales the $S_{(2j)}$ by a factor of $I_0^j$, and so sets the scales for each extracted order. The choice of $I_0$ has no other effect on the measured quantities. In the main text and below, we refer to intensities $I$ as having dimensions of nJ where we believe it is unambiguous; in such cases, the dimensionless intensities use $I_0 = 9$ nJ as a reference.

## S3 Random error

For the random error $\eta_k$ we determine the standard deviation in the raw 2DES data $S(I_k)$ by analyzing the low-intensity reference measurement. The data, collected for $M$ different values of $\tau$, is processed as described below in Section S8 to obtain the signal as a function of $\omega_\tau$. The goal is to find the contribution of random noise to $S^{nQ}(\tau = 0)$, which are the objects for which Eq. 14 allows us to estimate the systematic error. We construct that noise by assuming that the noise at each of the $M$ values of $\tau$ is independently chosen from a normal distribution with zero mean and standard deviation $\sigma_\tau$. We cannot measure $\sigma_\tau$ easily in our data, but we can easily measure the noise $\sigma_\omega$ in $\omega_\tau$ regions where no signal is expected. Here we show how the noise propagates into the frequency-domain signals and how measurement of the typical noise in a signal-free region of $\omega_\tau$ can give the desired expected noise on $S^{nQ}(\tau = 0)$. The result is Eq. S10, below.

In our signal analysis (Section S8), the data are reflected from positive to negative $\tau$ with the point at $\tau = 0$ not duplicated. We then have $M' = 2M - 1$ points along the $\tau$ axis. For this discussion, we use $S(\tau)$ to refer to the $\tau$-domain signal and $\tilde{S}(\omega_\tau)$ to refer to the frequency-domain signal.

We choose to put the $1/M'$ normalization on the forward discrete Fourier transform,

$$\tilde{S}(\omega_\tau, T, \omega_t, I) = \frac{1}{M'} \sum_\tau S(\tau, T, \omega_t, I) e^{i\omega_\tau \tau}. \tag{S5}$$

The inverse discrete Fourier transform is then

$$S(\tau, T, \omega_t, I) = \sum_{\omega_\tau} \tilde{S}(\omega_\tau, T, \omega_t, I) e^{-i\omega_\tau \tau}. \tag{S6}$$

We choose this convention so that the $\tau = 0$ point is a simple sum of the frequency-domain signal,

$$S(\tau = 0, T, \omega_t, I) = \sum_{\omega_\tau} \tilde{S}(\omega_\tau, T, \omega_t, I). \tag{S7}$$

Similarly, if we wish to obtain the window-integrated $n$Q signals at $\tau = 0$, we take the restricted sum

$$S^{nQ}(\tau = 0) = {\sum_{\omega_\tau}}' \tilde{S}(\omega_\tau), \tag{S8}$$

where the sum ranges depend on $n$.

After reflection, $S(\tau)$ is real and symmetric, which means that $\tilde{S}(\omega_\tau)$ is also real and symmetric. We can take the positive-frequency components as being independent normally distributed noise with standard deviation $\sigma_\omega$. To find the variance $\sigma_\omega^2$, we note that the variance of each $\eta_k$ is $\sigma_\tau^2$. Then from Eq. S5, the Fourier transform adds $\eta_0$ with variance $\sigma_\tau^2$ to two times $\cos(\omega_\tau \tau_i) \eta_i$ for each of the $M - 1$ terms with $\tau_i > 0$. Each of those terms has variance $4\cos^2(\omega_\tau \tau_i) \sigma_\tau^2$. The variance of the sum can be found by replacing $\cos^2$ with $1/2$. Dividing by $M'$ as in Eq. S5 gives

the root-mean-square (rms) noise amplitude at $\omega_\tau \neq 0$,

$$\sigma_\omega = \sigma_\tau \frac{\sqrt{1 + 2(M-1)}}{M'} = \frac{\sigma_\tau}{\sqrt{M'}} \tag{S9}$$

When summing over $L$ positive frequencies, as in Eq. S8, the rms magnitude of the sum is $\sqrt{L}\sigma_\omega$, since the frequency-domain signals are approximately independent real quantities with standard deviation $\sigma_\omega$. As a sanity check, we can sum over all frequencies to ensure that the noise in $\tau$ domain has standard deviation $\sigma_\tau$. The noise at $\omega_\tau = 0$ has variance $\sigma_\omega^2$. Each of the $M-1$ nonnegative frequencies is counted twice since the spectrum is symmetric and has variance $4\cos^2(\omega_\tau \tau)\sigma_\omega^2$. For $\tau \neq 0$, we can again approximate the $\cos^2$ as $1/2$, giving a total variance of $\sigma_\omega^2[1 + 2(M-1)] = \sigma_\omega^2 M'$, which is $\sigma_\tau^2$, as desired, according to Eq. S9. There are in fact some subtle correlations in the frequency-domain noise to ensure that the variance at $\tau = 0$ is also $\sigma_\tau^2$.

For the $n$Q signals at $\tau = 0$, we do not sum over all $M'$ frequency points. If the $n$Q window has $M'_{\text{window}}$ frequency points along $\omega_\tau$, the rms noise amplitude of $S^{n\text{Q}}(\tau = 0)$ is the sum of $M'_{\text{window}}$ independent values of amplitude $\sigma_\omega$, giving the $\sigma$ we use in the main text of

$$\sigma = \sqrt{M'_{\text{window}}} \sigma_\omega. \tag{S10}$$

This result shows how to take an observed value of noise in frequency domain $\sigma_\omega$ and get the effective noise relevant to the signals in Eq. 14 of the main text.

Our goal is to estimate the noise in $S^{n\text{Q}}(\tau = 0)$, which is found by summing only over positive frequency bins. We determine $\sigma_\omega$ by computing the rms average signal in the region where $0.6\,\text{eV} < \omega_\tau < 0.9\,\text{eV}$ and $1.5\,\text{eV} < \omega_t < 1.8\,\text{eV}$, which is in between the 0Q and 1Q signals' spectral positions, and we consider it to be free of signal and to contain only noise. We find by eye that the signal fluctuates around zero and thus is dominated by noise. Any systematic signal present is expected to come from the wings of the 0Q and 1Q signals, and therefore to vary with $\omega_t$. For each $\omega_t$, we therefore compute the mean signal along the $\omega_\tau$-axis in this region. We subtract the mean, which is still a function of $\omega_t$. We then compute the rms average of the signal over both $\omega_\tau$ and $\omega_t$ in this region, assuming that the noise is independent of $\omega_t$. We find $\sigma_\omega = 6.6 \times 10^{-8}$, so Eq. S10 gives $\sigma = 6 \times 10^{-7}$. We round this result to $5 \times 10^{-7}$. We have $M'_{\text{window}} = 85$, i.e., the number of pixels within the $n$Q regions ($\omega_0[n - 1/2, n + 1/2]$ for $\omega_0 = 1.59$ eV).

## S4  Connection between TA and 2DES signals

We here discuss in more detail the derivation of Eq. 12, reproduced here:

$$S^{\text{TA}}_{(2n)}(T, \omega_t) = S^{n\text{Q}}_{(2n)}(\tau = 0, T, \omega_t). \tag{S11}$$

We begin with the identity $S^{2\text{DES}}(\tau = 0, I) = S^{\text{TA}}(4I)$, where the factor of four originates in the different conventions for labeling intensity in the same experiment. In 2DES, there are commonly two identical pump pulses, each of

intensity $I$. When they overlap at $\tau = 0$ they form a single pump pulse of intensity $4I$. In contrast, a TA experiment has only a single pump pulse, whose intensity is labeled $I$. Thus in order to compare results, the factor of 4 is needed. Expanding $S^{\text{2DES}}(I) = \sum_n S^{\text{2DES}}_{(2n)} I^n$ and $S^{\text{TA}}(I) = \sum_n S^{\text{TA}}_{(2n)} I^n$, it follows that $S^{\text{2DES}}_{(2n)}(\tau = 0) = 4^n S^{\text{TA}}_{(2n)}$. By the properties of the Fourier transform, we have that

$$S^{\text{2DES}}_{(2n)}(\tau = 0) = \int d\omega_\tau S^{\text{2DES}}_{(2n)}(\omega_\tau), \tag{S12}$$

which is equivalent to

$$S^{\text{2DES}}_{(2n)}(\tau = 0) = \sum_{r=-\infty}^{\infty} S^{r\text{Q}}_{(2n)}(\tau = 0) \tag{S13}$$

which includes $r < 0$ signals centered at $\omega_\tau < 0$. The integral and the sum are equivalent, so long as the windows used to integrate the $r$Q signals are contiguous and span the entire $\omega_\tau$ axis. These $n$Q with $n < 0$ signals obey $\left[S^{-|n|\text{Q}}(-\omega_\tau)\right]^* = S^{|n|\text{Q}}(\omega_\tau)$, as required given that the time-domain signal in pump-probe geometry is real. The imaginary parts cancel upon full integration over $\omega_\tau$, or equivalently summing over all $n$Q. Then at each order we have

$$\sum_{r=-\infty}^{\infty} S^{r\text{Q}}_{(2n)}(\tau = 0) = S^{n\text{Q}}_{(2n)}(\tau = 0) \sum_{r=-n}^{n} \binom{2n}{n - |r|}, \tag{S14}$$

where the left-hand side comes from Eq. 11. We prove below that $\sum_{r=-n}^{n} \binom{2n}{n-|r|} = 4^n$, and we conclude that

$$\sum_{r=-\infty}^{\infty} S^{r\text{Q}}_{(2n)}(\tau = 0) = 4^n S^{n\text{Q}}_{(2n)}(\tau = 0) \tag{S15}$$

and so at each order we have

$$S^{n\text{Q}}_{(2n)}(\tau = 0) = S^{\text{TA}}_{(2n)}. \tag{S16}$$

We now prove that $\sum_{r=-n}^{n} \binom{2n}{n-|r|} = 4^n$ using the binomial theorem. We consider $(x + y)^{2n}$, which we expand as

$$(x + y)^{2n} = \sum_{k=0}^{2n} \binom{2n}{k} x^j y^{2n-j} = 2 \sum_{k=0}^{n} \binom{2n}{k} x^j y^{2n-j} - \binom{2n}{n} \tag{S17}$$

where in the last equality we subtract the double-counted term when summing only to $n$ instead of $2n$. Then let $x = y = 1$ and let $r = n - k$, so

$$4^n = 2 \sum_{r=0}^{n} \binom{2n}{n - r} - \binom{2n}{n}. \tag{S18}$$

Note that for any function $f_r$, we have $2\sum_{r=0}^n f_r = f_0 + \sum_{r=-n}^n f_{|r|}$. We then have

$$4^n = \sum_{r=-n}^{n} \binom{2n}{n - |r|}. \tag{S19}$$

## S5 Analytical saturation model for 2DES signals

Here we show that if the intensity-dependent TA signal follows an exponential saturation model as in Eq. 13, $S^{\text{TA}}(T, \omega_t, I) = -S_{\max}(T, \omega_t)\left(1 - e^{-I/I_{\text{sat}}(T,\omega_t)}\right)$, then the $n$Q signals at $\tau = 0$ obey

$$S^{n\text{Q}}(\tau = 0, I) = \sum_{r=n}^{\infty} S_{(2r)}^{n\text{Q}} I^r = S_{\max} \begin{cases} e^{-2I/I_{\text{sat}}} \mathcal{I}_0(2I/I_{\text{sat}}) - 1 & \text{for } n = 0, \\ (-1)^r e^{-2I/I_{\text{sat}}} \mathcal{I}_r(2I/I_{\text{sat}}) & \text{for } n \geq 1, \end{cases} \quad (S20)$$

which is Eq. 14. We begin by taking the Taylor series of $S^{\text{TA}}$ to get

$$S^{\text{TA}}(I) = S_{\max} \sum_{j=1}^{\infty} \frac{(-1)^j}{j!} \left(\frac{I}{I_{\text{sat}}}\right). \quad (S21)$$

From this form, we immediately have

$$S_{(2j)}^{\text{TA}} = S_{\max} \frac{(-1)^j}{j! I_{\text{sat}}^j}. \quad (S22)$$

Then using Eqs. 11 and 12, we have $S_{(2n)}^{r\text{Q}}(\tau = 0) = \binom{2n}{n-r} S_{(2n)}^{\text{TA}}$, which gives

$$S_{(2n)}^{r\text{Q}}(\tau = 0) = \binom{2n}{n-r} S_{\max} \frac{(-1)^n}{n! I_{\text{sat}}^j} \quad (S23)$$

and therefore

$$S^{r\text{Q}}(\tau = 0, I) = S_{\max} \sum_{n=1}^{\infty} \binom{2n}{n-r} \frac{(-1)^n}{n!} \left(\frac{I}{I_{\text{sat}}}\right)^n. \quad (S24)$$

This series can be summed in closed form, and Mathematica gives Eq. S20.

We note that this same procedure can be completed for any model $S^{\text{TA}}(I)$, so long as it has a Taylor series with a finite radius of convergence, though a closed form may be impossible to find. For example, we demonstrate the same procedure for the case called "saturable absorption" in Section S6.2.

## S6 Choosing optimal intensities

The main text presents the theoretical background for estimating random and systematic (contamination) errors for 2DES in the weak-probe limit, allowing determination of optimal intensities. Here, we show the details of the experimental procedure.

### S6.1 Systematic error: exponential saturation

For the systematic error, we find the parameters $I_{\text{sat}}$ and $S_{\max}$ from the saturation form of the TA spectra. We performed TA measurements for 101 linearly spaced excitation-pulse intensities between 0.27 nJ and 276 nJ at a population time $T = 2$ ps with identical pulse envelope shapes, with intensity controlled by the pulse shaper and all other experimental parameters kept unchanged. Figure S2a shows one such saturation curve, at the peak linear-

absorption energy $\hbar\omega_t = A = 1.58$ eV. A fit (grey line) to

$$S_{\text{TA}}^{\text{exponential}}(I) = -S_{\max}(1 - e^{-I/I_{\text{sat}}}), \tag{S25}$$

which is Eq. 13, is shown. Figures S2b and c show $S_{\max}$ and $I_{\text{sat}}$, respectively, extracted from such a fit at each $\omega_t$. We find $I_{\text{sat}}$ ranging from 27 to 106 nJ and $S_{\max}$ ranging from 2.0 to 55.6 $\Delta$mOD within the spectral range between 1.50 eV and 1.76 eV. Our goal is to choose $S_{\max}$ and $I_{\text{sat}}$ that are typical for the spectrum and/or at an $\omega_t$ important for the 2D spectra. In the main text, we use the values at $A$, the peak in the absorption spectrum. The standard errors in the extracted parameters are smaller than 1.5 nJ and 0.2 $\Delta$mOD for $I_{\text{sat}}$ and $S_{\max}$, respectively, consistent with the high quality of the fits.

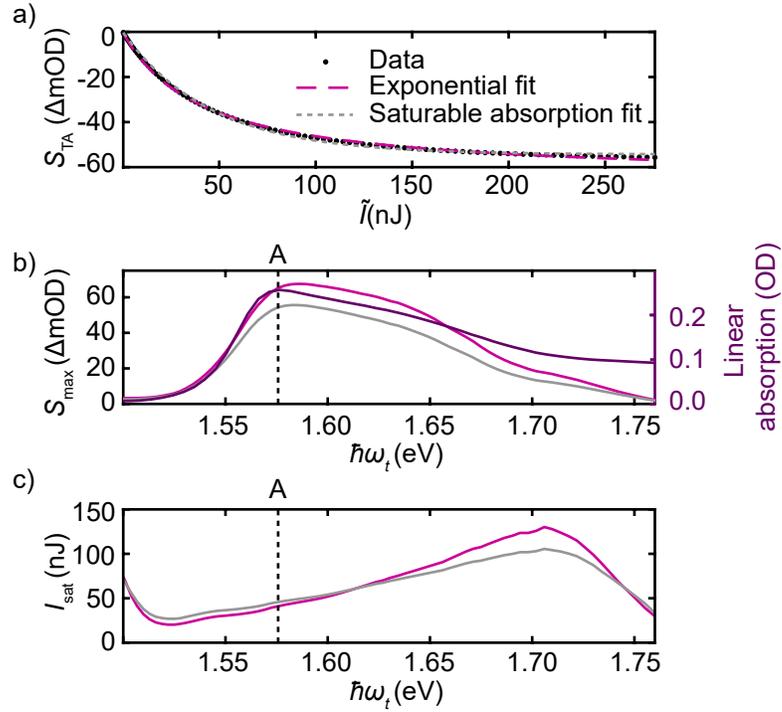

Figure S2: Systematic error estimation using power-dependent TA data. (a) Measured $S_{\text{TA}}(I)$ at $\hbar\omega_t$=A= 1.58 eV and fit to Eq. 13 (grey) and Eq. S26 (magenta). (b) Fit parameter $S_{\max}$ for all $\omega_t$ and linear absorption spectrum. (c) Fit parameter $I_{\text{sat}}$ for all $\omega_t$. The dashed lines at 1.58 eV indicate the frequency shown in a, which is also used for the optimal intensity calculation.

### S6.2 Systematic error: saturable absorption

While Figure S2a (grey line) shows the good quality of the exponential saturation form for fitting $S_{\text{TA}}(I)$, we also consider saturable absorption, defined by

$$S_{\text{TA}}^{\text{sat}}(I) = -S_{\max}\frac{I}{I + I_{\text{sat}}}. \tag{S26}$$

Just like $S_{\text{TA}}^{\text{exponential}}(I)$, this form has a known Taylor series, $S_{\text{TA}}^{\text{sat}}(I) = -S_{\max} \sum_{n=1}^{\infty} \left(\frac{-I}{I_{\text{sat}}}\right)^n$. This power series is convergent only for $I < I_{\text{sat}}$, unlike in Eq. S25. We now show how the analysis of the main text for exponential absorption can be straightforwardly adapted to $S_{\text{TA}}^{\text{sat}}$ to find the optimal intensities for a 2DES experiment focused on a particular $n$Q spectral region. Then we show how well our TA results fit Eq. S26 and the optimal intensities for 2DES suggested by those fits. Finally, we compare both saturation models with the extracted 2D data.

As in Eq. S23, we connect the TA response orders $S_{\text{TA},(n)}^{\text{sat}} = -S_{\max}(-1/I_{\text{sat}})^n$ to the $r$Q response orders,

$$S_{(2n)}^{r\text{Q}}(\tau = 0) = -\overbrace{\frac{S_{\max}}{I_{\text{sat}}^n}(-1)^n}^{S_{\text{TA},(n)}^{\text{sat}}} \binom{2n}{n-r}. \tag{S27}$$

Then letting $\tilde{I} \equiv I/I_{\text{sat}}$,

$$S^{r\text{Q}}(\tilde{I}, \tau = 0) = \sum_n S_{(2n)}^{r\text{Q}}(\tau = 0) I^n = S_{\max} \sum_n (-1)^{n+1} \binom{2n}{n-r} \tilde{I}^n = S_{\max}(-4\tilde{I})^r \left(\frac{1}{\left(\sqrt{4\tilde{I}+1}+1\right)^{2r} \sqrt{4\tilde{I}+1}} - \delta_{r0}\right), \tag{S28}$$

where $\delta_{ij}$ is the Kronecker delta function and the last identity comes from Mathematica. For saturable absorption, this new function replaces the modified Bessel functions that we found for exponential saturation. Note that $S^{r\text{Q}}(\tilde{I}, \tau = 0)$ has a power series that is convergent only for $\tilde{I} < 1/4$.

For a given set of pump intensities $\{\tilde{I}_k\}$, we perform the Vandermonde extraction of $\hat{S}_{(2n)}^{r\text{Q}}$ using Eq. 5 and define the systematic error as before, by comparing the Vandermonde-extracted $S_{(2n)}^{r\text{Q}}(\tau = 0)$ to the exact values in Eq. S27. We use the same noise $\sigma$ as in Section S3. The updated optimal errors and pulse intensities are shown in Table S1, which is equivalent to Table 1 in the main text.

Table S1: Equivalent to Table 1 but using saturable absorption for the TA spectra instead of exponential saturation. Errors in $M = 3$ orders in 1Q and 2Q spectra at $\tau = 0$ for several choices of number of intensities $N$. We use $\tilde{\sigma} = 7.7 \times 10^{-6}$, with $\sigma$ as in Section S3 and $S_{\max} = 0.065 \ \Delta\text{OD}$.

| $N$ | 1Q error | Optimal $I/I_{\text{sat}}$ for 1Q | 2Q error | Optimal $I/I_{\text{sat}}$ for 2Q |
|---|---|---|---|---|
| 3 | 0.15 | 0.0067, 0.027, 0.043 | 0.22 | 0.0082, 0.033, 0.053 |
| 4 | 0.08 | 0.0092, 0.040, 0.074, 0.10 | 0.14 | 0.011, 0.047, 0.086, 0.11 |
| 5 | 0.06 | 0.011, 0.050, 0.097, 0.14, 0.17 | 0.11 | 0.012, 0.057, 0.11, 0.16, 0.19 |
| 6 | 0.05 | 0.012, 0.056, 0.11, 0.18, 0.23, 0.25 | 0.09 | 0.014, 0.063, 0.13, 0.20, 0.25, 0.28 |

For a system obeying saturable absorption, contamination errors are larger than for a system with exponential saturation, so the optimal intensities for order extraction are lower. Similarly, with fixed saturated noise-to-signal ratio $\tilde{\sigma}$, the minimum possible errors in order extraction are higher. These phenomena also mean the optimal intensity selection is less affected by $I_{\max}$. In our case, the optimal intensities do not exceed $I_{\max}$ until $N = 8$. We note that even though the power series for $S^{r\text{Q}}(I, \tau = 0)$ are divergent for $I > I_{\text{sat}}/4$, the optimal intensities for order extraction can exceed $I_{\text{sat}}/4$. Using the experimental intensities, which were chosen using the exponential saturation form, we find expected 1Q and 2Q errors of 0.29 and 0.39, respectively. This error is largely due to contamination, since the experimental intensities exceed $I_{\text{sat}}$.

Figure S2 shows that the intensity-dependent TA data are well fit by the exponential saturation form, Eq. 13, but the saturable form of Eq. S26 fits just as well. This observation raises the question whether the intensities used for the order extraction were, in fact, optimal. The 2D data must be consistent with the TA data when integrating over $\omega_\tau$. In particular, the two saturation forms make different predictions for $S^{rQ}_{(2n)}(\tau = 0)$,

$$S^{rQ}_{(2n),\text{exp}}(\tau = 0) = \frac{S_{\max}}{I^n_{\text{sat}}} \binom{2n}{n-r} \frac{(-1)^{n+1}}{n!}, \tag{S29}$$

$$S^{rQ}_{(2n),\text{sat}}(\tau = 0) = \frac{S_{\max}}{I^n_{\text{sat}}} \binom{2n}{n-r} (-1)^{n+1}. \tag{S30}$$

Equations S29 and S30 do not include the effects of systematic errors. Therefore, instead of comparing our extracted orders to these Taylor series expansions, we compare the experimental extracted orders to the theoretical extractions $\hat{S}^{rQ}_{(2n),\text{exp}}$ and $\hat{S}^{rQ}_{(2n),\text{sat}}$ that are obtained by evaluating the respective intensity-dependent models, equations S20 and S28, at the experimental pulse intensities, and applying Eq. 5 of the main text. Using the extracted $S^{rQ}_{(2n)}(\omega_\tau, T, \omega_t)$, we window-integrate over $\omega_\tau$ to obtain $S^{rQ}_n(\tau = 0, T, \omega_t)$. We compare the experimental results to $\hat{S}^{rQ}_{(2n),\text{exp}}$ and $\hat{S}^{rQ}_{(2n),\text{sat}}$ with fixed $r$ and varying $n$. Figure S3 shows the $\hat{S}^{rQ}_{(2n)}(\tau = 0, T, \omega_t)$ at $T = 2$ ps and $I_0 = 9$ nJ for $\omega_t = 1.57$ eV and 1.58 eV along with $\hat{S}^{rQ}_{(2n),\text{exp}}(\tau = 0)$ and $\hat{S}^{rQ}_{(2n),\text{sat}}(\tau = 0)$, where $I_{\text{sat}}$ is determined by the fits shown in Figure S2. Note that $S^{2Q}_{(2)} = 0$, so the model signals there are entirely due to contamination error.

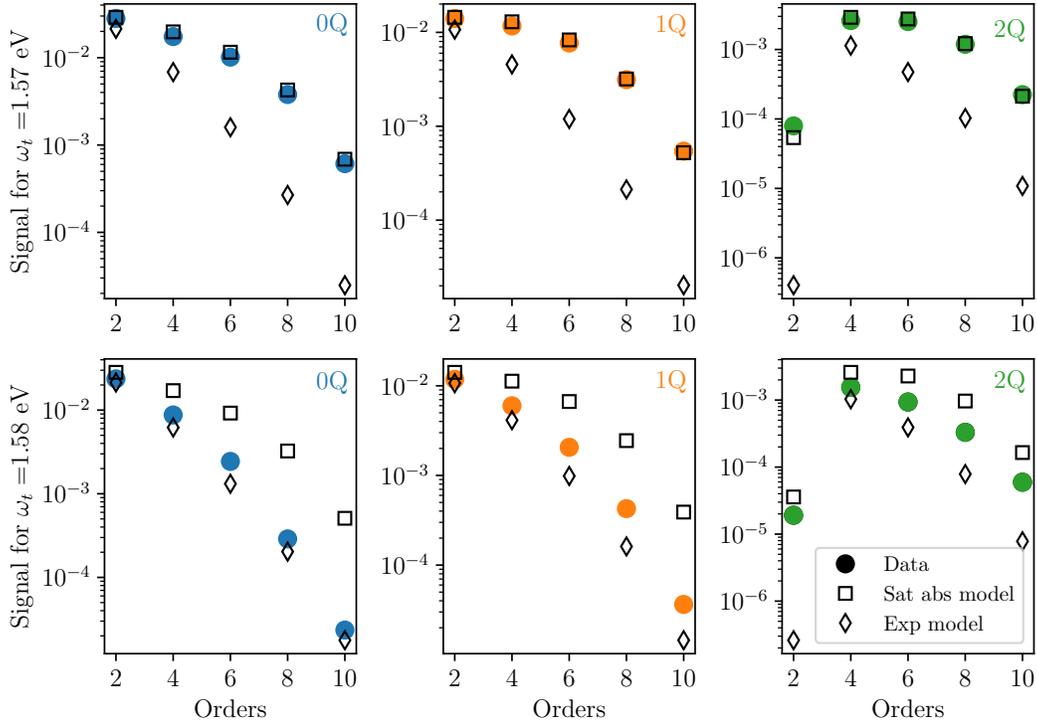

Figure S3: Comparison of saturation models. Integrated spectral regions of the 0Q-2Q signals ($\omega_0[r - 1/2, r + 1/2]$ for $\omega_0 = 1.59$ eV) at $T = 2$ ps and $I_0 = 9$ nJ and predictions from two saturation models: Exponential saturation model ($\hat{S}^{rQ}_{(2n),\text{exp}}(\tau = 0)$, diamonds) described by Eq. S29 and saturable absorption model ($\hat{S}^{rQ}_{(2n),\text{sat}}(\tau = 0)$, squares), described by Eq. S30.

Figure S3 shows that the experimental data agree better with the saturable absorption model at $\omega_t = 1.57$ eV, while the data seem to agree better with the exponential saturation model at $\omega_t = 1.58$ eV. In some spectral regions neither model fits well. We do not believe that either saturation form is precisely accurate but rather that they serve as guides to aid in the selection of optimal intensities. We do not anticipate perfect agreement of extracted orders with the predictions of either model. We conclude that either model can be used for this set of data.

### S6.3 Optimization for TA spectroscopy

We consider now the optimization of intensities for extraction of $M = 3$ orders in TA spectroscopy, rather than for $n$Q 2DES. The process of estimating the contamination error is the same as described in the main text, but we use the saturation form for TA spectra (e.g., Eq. 13 or Eq. S26) in place of the saturation form for $n$Q spectra (Eq. 14 or Eq. S28). We illustrate here for the exponential saturation case, Eq. 13. For the random noise distribution, we use $\sigma_\tau = \sqrt{M'}\sigma_\omega = 1.6 \times 10^{-6}$ with $\sigma_\omega$ as determined in Section S3, which gives $\tilde{\sigma} = 3.0 \times 10^{-5}$. Then the optimal errors and intensities for the TA measurement are shown in Table S2. These TA pump intensities are similar to those obtained at $\tau = 0$ from the 1Q 2D pump intensities in Table 1, which produce a TA pulse with four times larger intensity.

Table S2: Equivalent to Table 1 but for extraction of orders of the TA signal rather than of $n$Q signals, using the exponential saturation form of the TA spectra. Errors in $M = 3$ orders for several choices of number of intensities $N$.

| $N$ | TA error | Optimal $I/I_{\text{sat}}$ |
|---|---|---|
| 3 | 0.094 | 0.055, 0.22, 0.35 |
| 4 | 0.036 | 0.085, 0.37, 0.67, 0.86 |
| 5 | 0.020 | 0.11, 0.49, 0.96, 1.4, 1.6 |
| 6 | 0.013 | 0.13, 0.60, 1.2, 1.8, 2.3, 2.6 |

## S7 Experimental setup

The TA measurements are taken with a partially non-collinear pump-probe setup. The output of a commercial Yb-laser (PHAROS, Light Conversion) is spectrally broadened with a commercial non-collinear optical parametric amplifier (ORPHEUS, Light Conversion) and yields pulses with a spectral range from 550 nm to 840 nm at a repetition rate of 50 kHz. The amplified output is then split (178892, Layertec GMbH), with 90% of the intensity in the pump beam and the remaining 10% in the probe beam. The probe pulse is compressed by two fused silica prisms and time delayed by a mechanical stage (M-IMS1000LM-S, Newport) to adjust the population time $T$. The pump pulse is pre-compressed by two BK7 prisms and then compressed by an acousto-optic modulator (AOM) pulse shaper (Quickshape, PhaseTech) that also generates coherent double pulses with variable inter-pulse delays for the 2D measurements. The pump pulse has a duration of $\tau_p = 14$ fs (intensity FWHM) after compression, characterized via a collinear frequency resolved optical gating (FROG) setup. The spectral region from 550 nm to 660 nm is blocked in the prism compressor setups for both beams. After compression, both beams are focused in the sample position and have beam radii of $r_x = 43$ µm and $r_y = 38$ µm for the probe and $r_x = 67$ µm and $r_y = 172$ µm for the pump

beam at a limit of $e^{-1}$ of the maximum amplitude. The pump beam is blocked after passing through the sample, and the probe beam is detected on a shot-to-shot basis with a spectrometer (Spektrometer Acton 2156, Princeton Instruments) and line camera (HS-Kamera Serie3030, Entwicklungsbüro Stresing). The data are taken and evaluated with custom LabView 2021 and Matlab R2023b scripts. For the extraction of response orders of the 2D data, a total of six 2D measurements at different pump powers were taken. Each 2D spectrum was recorded while scanning the time delay $\tau$ between two pump pulses over $M = 299$ steps with a step size of 0.37 fs. The population time was set to $T = 2$ ps. First the full 2D spectrum was measured before the intensity was changed. The pump-pulse energies $E_p$ were set to 15 nJ, 14 nJ, 10 nJ, 5.5 nJ, 1.2 nJ, and 0.16 nJ, chosen according to the optimal intensity procedure described in the main text and in Section S6. The pump-pulse energies were determined by measuring the pulse energy of both excitation pulses at $\tau = 0$ fs and dividing them by four to obtain the intensity of a single pump pulse. The first five 2D spectra were used for the extraction of the different orders, while the measurement at the lowest pump energy is used as low-intensity reference. In this work we refer to the intensities by stating the pulse energies as we do not vary the pump pulse envelope shape or the spot size. The maximum pulse intensities are given within the approximation of a Gaussian beam by

$$I_{\text{pump}} = \frac{4E_p}{\tau_p r_x r_y}\sqrt{\frac{\ln 2}{\pi^3}}, \tag{S31}$$

which yields the peak pulse intensities listed in Table S3.

**Table S3: Pulse energies and pulse intensities of a single pump pulse in 2D experiments.**

| $E_p$ (nJ) | $I_{\text{pump}}$ (GWcm$^{-2}$) |
|---|---|
| 15 | 5.6 |
| 14 | 5.1 |
| 10 | 3.7 |
| 5.5 | 2.0 |
| 1.2 | 0.46 |
| 0.16 | 0.058 |

## S8  Raw data and data processing

Before extracting the higher-order spectra discussed in the main manuscript, we apply the following data processing procedure. First, we calculate the transient absorption signal $S(\tau, T, \omega_t, I) = -\log_{10} \frac{S_{\text{pumped}}(\tau, T, \omega_t, I)}{S_{\text{unpumped}}(\tau, T, \omega_t, I)}$. To Fourier transform with respect to $\tau$, for each dataset $s(\tau) = S(\tau, T, \omega_t, I)$, we first reflect the signal to $-\tau$, without duplicating the $\tau = 0$ point and then perform the discrete Fourier transform, using Eq. S5.

The squaraine polymer linear absorption (Figure 4a) signal reveals two main peaks at 1.58 eV ($A$) and 1.87 eV originating from the absorption of the SQA and SQB sites. The J-type coupling leads to a further splitting of the peaks and the enhancement of the peak $A$ at 1.58 eV in comparison to the separated monomer units.[1] The pump spectrum covers only the spectral region from the low-energy exciton peak at 1.58 eV ($A$) to the vibrational mode 1.73 eV ($D$), but not the high-energy exciton state at 1.87 eV.

The 1Q (around $\omega_0$) and 2Q (around $2\omega_0$) regions of the raw data, used for extracting up to the $S_{(10)}$ spectra,

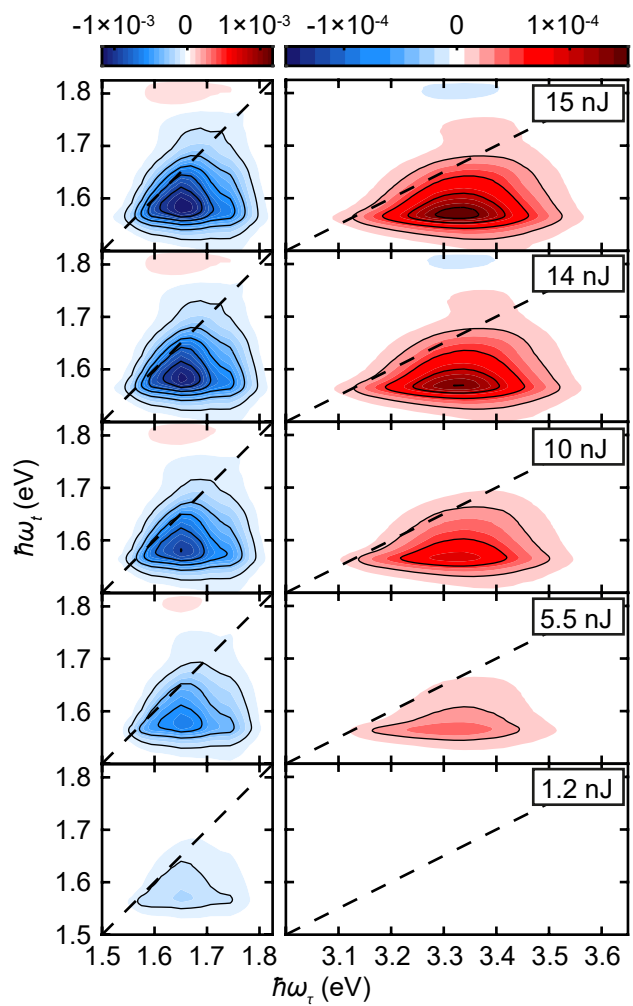

Figure S4: Raw 2D data at the 1Q and 2Q position taken at five different excitation intensities.

are shown in Figure S4. All 1Q and 2Q spectra display an intense signal slightly shifted below the diagonal, due to energy transfer taking place during the $T = 2$ ps population time.[2,3] The 1Q spectra are dominated by a negative feature, whereas the 2Q spectra exhibit a positive main feature. Additionally, a feature emerges at the blue edge of the $\omega_t$ axis (1.81 eV) in both $n$Q spectra, with its sign inverted relative to the corresponding main feature. With increasing excitation intensity, the absolute 2Q signal is enhanced compared to the 1Q signal.

Along $\omega_\tau$, the 1Q spectra display three distinct peaks, which correspond to the peaks in the absorption spectrum $A$, $C$, and $D$ (Figure 4a). The signal maximum of all 1Q spectra lies at the $C$ peak (1.65 eV) although the extracted order spectra exhibit a shift towards the redshifted peak $A$ (Figure 4b). Moreover, the 2Q spectra exhibit an elongated shape without clearly separable peaks. The 2Q absolute signal maximum is blueshifted along $\omega_\tau$ compared to twice the absolute signal maximum of the 1Q spectrum, agreeing with the predictions of Bubilaitis and Abramavičius for a model J-aggregate.[4]

## S9   Comparison of lowest order with low-intensity reference measurement

To verify the success of the extraction, we analyzed the integrated $n$Q signal of $S_{(2)}$ and $S_{(4)}$ (Figure 4b). For one further verification of the extraction of the lowest-order signal, we compare our extracted lowest order signal, $S_{(2)}$ (Figure S5a), with a low-intensity reference measurement, $S_L$ (Figure S5b), detected at 0.16 nJ, where we assume negligible higher-order contributions. To facilitate the direct comparison, we scale $S_L$ to the reference power of the extraction $I_0 = 9$ nJ (Figure S5). The difference signal (Figure S5c) and the comparison of the integrated 1Q signals (Figure S5d) confirm the correct extraction of the pure third order.

## S10   Visibility enhancement of cross peaks in higher-order signals

In the experimental data we see that there is a cross peak at the $(\omega_\tau, \omega_t) = (\omega_A, \omega_B)$ position that is visible in $S_{(4)}^{1Q}$ but does not appear in $S_{(2)}^{1Q}$. In this section we show that for a simple toy model the upper cross peak in the 1Q signal is frequently enhanced in the $S_{(4)}$ signal as compared with $S_{(2)}$. The high-order extraction thus gives visibility into this process even when it is not visible in the standard leading-order 2DES. The experimental data are for a polymer with on average 18 dimer subunits. For this theoretical discussion, we consider a dimer of two-level systems, which is the simplest possible model of a polymer. The fact that we see enhancement of the cross peak in $S_{(4)}$ as compared with $S_{(2)}$ for a wide range of parameters for such a simple model leads us to conclude that such effects can occur, and are perhaps even highly likely to occur, in polymer systems.

We describe our dimer with the Hamiltonian

$$H = \begin{pmatrix} \omega_g & 0 & 0 & 0 \\ 0 & \omega_a & J & 0 \\ 0 & J & \omega_b & 0 \\ 0 & 0 & 0 & \omega_f \end{pmatrix}, \tag{S32}$$

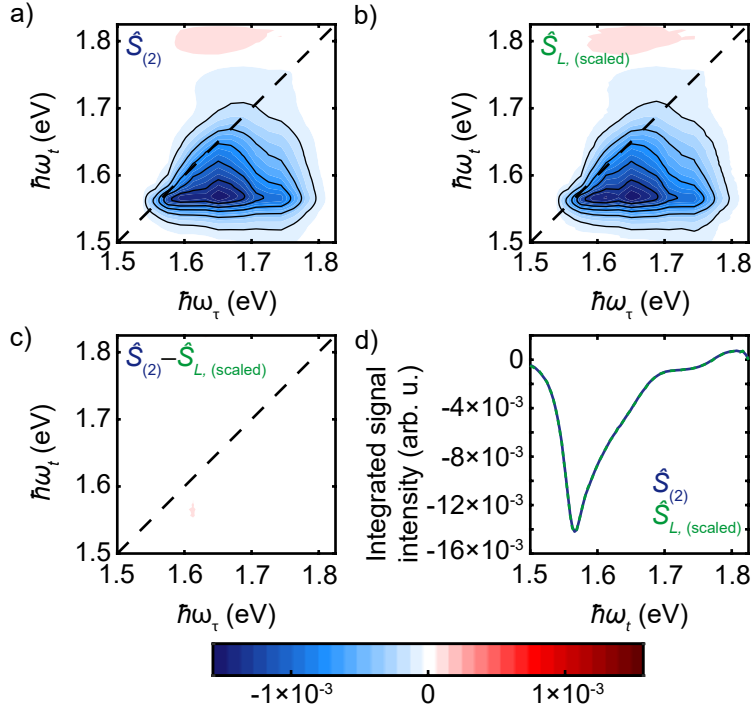

Figure S5: Lowest-order signal 1Q spectra. a) Extracted $\hat{S}_{(2)}$, b) Low-intensity reference measurement $S_{L,\text{scaled}}$ scaled to $I_0$ (9 nJ). c) Scaled low-intensity measurement subtracted from the extracted $\hat{S}_{(2)}$. d) $\hat{S}_{(2)}$ and $S_{L,\text{scaled}}$ integrated along $\omega_\tau$ 1Q ($\omega_0[0.5, 1.5]$ for $\omega_0 = 1.59$ eV).

where $\hbar = 1$ and we take $\omega_g = 0$ and $\omega_f = \omega_a + \omega_b$, where there are four basis states $|g\rangle, |a\rangle, |b\rangle, |f\rangle$. The Hamiltonian can be diagonalized to find the eigenvalues $0, \omega_\alpha, \omega_\beta, (\omega_a + \omega_b)$, where

$$\omega_\alpha = \frac{1}{2}\left(\omega_a + \omega_b - \sqrt{(\omega_a - \omega_b)^2 + 4J^2}\right), \tag{S33}$$

$$\omega_\beta = \frac{1}{2}\left(\omega_a + \omega_b + \sqrt{(\omega_a - \omega_b)^2 + 4J^2}\right),$$

corresponding to eigenstates

$$|\alpha\rangle = \frac{1}{\sqrt{N}}\left[-\left(x + \sqrt{x^2 + 1}\right)|a\rangle + |b\rangle\right], \tag{S34}$$

$$|\beta\rangle = \frac{1}{\sqrt{N}}\left[|a\rangle + \left(x + \sqrt{x^2 + 1}\right)|b\rangle\right],$$

where $N = 2\left(x^2 + 1 + x\sqrt{x^2 + 1}\right)$ with $x = \frac{\omega_b - \omega_a}{2J}$. We define the splitting between the singly excited eigenvalues

$$\omega_{\beta\alpha} = \omega_\beta - \omega_\alpha$$
$$= \sqrt{(\omega_a - \omega_b)^2 + 4J^2} \tag{S35}$$

and the average singly excited eigenvalue is

$$\omega_{\text{ave}} = \frac{\omega_\alpha + \omega_\beta}{2} = \frac{\omega_a + \omega_b}{2}. \tag{S36}$$

With this model, we explore the cross peak at $(\omega_\tau, \omega_t) = (\omega_A, \omega_B)$, so we fix $\omega_\alpha = \omega_A$ and $\omega_\beta = \omega_B$. With these two values set, there is a single free parameter in the Hamiltonian, $J$, which can vary from 0 to $\omega_{\beta\alpha}/2$. For $J = 0$, there cannot be cross peaks in a 2DES signal, since the two subsystems are uncoupled. As $J$ increases, we expect that the cross peaks become more visible.

The dimer's dipole operator in the $|g\rangle, |a\rangle, |b\rangle, |f\rangle$ basis is

$$\mu = \begin{pmatrix} 0 & \mu_{ag} & \mu_{bg} & 0 \\ \mu_{ag} & 0 & 0 & \mu_{fa} \\ \mu_{bg} & 0 & 0 & \mu_{fb} \\ 0 & \mu_{fa} & \mu_{fb} & 0 \end{pmatrix}, \tag{S37}$$

where $\mu_{ij} = \langle i | \mu | j \rangle$, and we take $\mu_{fa} = \mu_{bg}$, $\mu_{fb} = \mu_{ag}$. Transforming to the eigenbasis, we have

$$\mu_{\alpha g} = \frac{1}{\sqrt{N}} \left[ -\left(x + \sqrt{x^2 + 1}\right) \mu_{ag} + \mu_{bg} \right], \tag{S38}$$

$$\mu_{\beta g} = \frac{1}{\sqrt{N}} \left[ \mu_{ag} + \left(x + \sqrt{x^2 + 1}\right) \mu_{bg} \right].$$

Let us assume that we have a linear absorption spectrum with two well-resolved peaks. In such a case, $\omega_\alpha$, $\omega_\beta$, $\mu_{\alpha g}$, and $\mu_{\beta g}$ are all known. Given these constraints, we can explore all possible dimer models that are consistent with this linear absorption spectrum, where there is a single free parameter, $J$, which can vary from 0 to $\omega_{\beta\alpha}/2$.

In the case of the experimental data, we know the splitting $\omega_{\beta\alpha}$ thanks to the cross peak visible in $S^{1Q}_{(4)}$, but we do not know the values of $\mu_{\alpha g}$ and $\mu_{\beta g}$, since the peaks overlap in the linear absorption spectrum. We will therefore consider three values of $\mu_r = \mu_{\alpha g}/\mu_{\beta g}$. We consider $\mu_r^2 = 0.5, 1, 2$. See the example linear absorption spectra for each of these three cases in Figure S6a where the spectra are normalized such that $\mu_{\alpha g}^2 + \mu_{\beta g}^2 = 1$. The ratio of the input dipole moments is

$$\frac{\mu_{bg}}{\mu_{ag}} = \frac{\mu_r + x + \sqrt{x^2 + 1}}{1 - \mu_r x - \mu_r \sqrt{x^2 + 1}}. \tag{S39}$$

Using the Ultrafast Spectroscopy Suite (UFSS)[5,6], we simulate the signals $S^{1Q}_{(2)}$ and $S^{1Q}_{(4)}$ due to impulsive pulses for population times $T\omega_{\beta\alpha} = 0.1, \pi/2, \pi$. We find the population time with the largest-magnitude cross peak, which we label $T'$. We then divide the cross-peak signal, $S_{(2n)}(\omega_A, T', \omega_B)$, by the absolute maximum signal $S^{\max}_{(2n)}(T') \equiv \max_{\omega_\tau, \omega_t} |S_{(2n)}(\omega_\tau, T', \omega_t)|$, to define the cross peak visibility as

$$\eta_{(2n)} = \frac{S_{(2n)}(\omega_A, T', \omega_B)}{S^{\max}_{(2n)}(T')}. \tag{S40}$$

We plot $\eta_{(2)}$ and $\eta_{(4)}$ in Figure S6b and the visibility enhancement $\eta_{(4)}/\eta_{(2)}$ in Figure S6c. The visibility enhancement

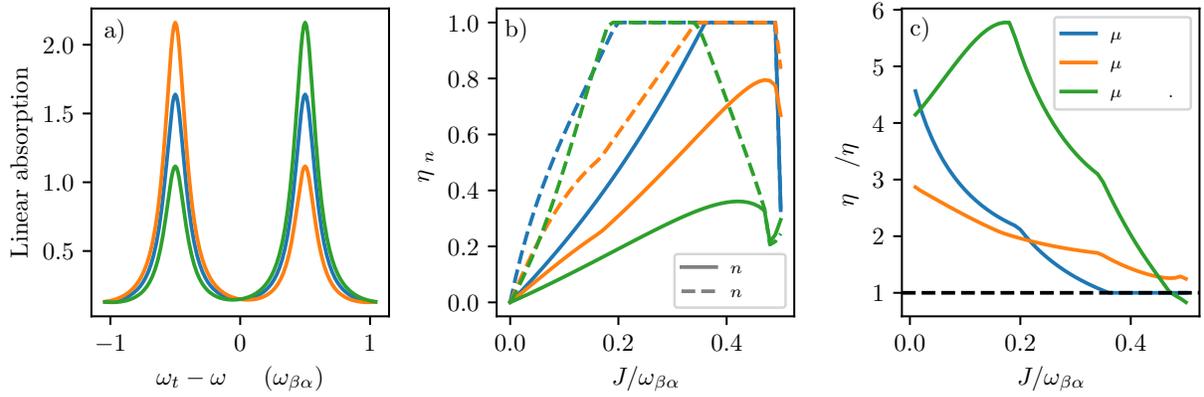

Figure S6: Dimer model. (a) Linear absorption for three choices of $\mu_r$. (b) Visibility $\eta_{(n)}$ of $(\omega_\alpha, \omega_\beta)$ cross peak for $n = 2, 4$ and the three choices of $\mu_r$ shown in panel a. (c) Relative visibility enhancement showing that the cross peak is almost always more visible in $S_{(4)}$ than it is in $S_{(2)}$.

is nearly always greater than 1, except for $J$ near its maximum value. In addition, the visibility enhancement is largest when visibility is small. This enhancement demonstrates the utility of measuring $S_{(4)}$, as it is a useful tool in uncovering cross peaks that might otherwise be hidden.

All of the calculations in this section are for a closed system and ignore the effects of a bath. The linear absorption spectra shown in Figure S6a are shown with a phenomenological linewidth of $\Gamma = 0.1\omega_{\beta\alpha}$, added after the simulations by multiplying the time-domain signal by $e^{-\Gamma t}$. However, the visibilities $\eta_{(2n)}$ are calculated by using a linewidth of $\Gamma = 0.01\omega_{\beta\alpha}$, This small linewidth ensures that the cross-peak position has negligible amplitude bleeding over from the diagonal peaks. When the linewidths are larger, the wings of the diagonal peaks can mask the cross peak. In this model, the upper-left cross peak in $S_{(2)}$ is always opposite in sign to the other three peaks for $J < 0.28$, and so the wings of the diagonal peaks can cancel with the cross peak, leaving little or no signal at that spectral location. We show two examples where the upper-left cross peak is not clearly visible in $S_{(2)}$ due to competition between a small but non-zero cross-peak signal and the wings of the diagonal peaks. Figures S7a and c show $S_{(2)}$ and $S_{(4)}$, respectively, for $J = 0.017\omega_{\beta\alpha}$ and the same linewidth used for Figure S6a. Figures S7b and d show $S_{(2)}$ and $S_{(4)}$, respectively, for $J = 0.1\omega_{\beta\alpha}$ and a Lorentzian linewidth of $\Gamma = 0.5$. Even though they come from a simple coupled-dimer model, panels b and d look qualitatively similar to the experimental results in Fig. 4b.

We further note that the cross peaks in a dimer model represent signals originating from transitions between both the ground state and the singly excited states, as well as signals originating from transitions between singly excited and doubly excited states. We have given all of these the same linewidth, but a more realistic model would have different linewidths, which would cause the cross peak to have more complicated shapes. However, we expect the qualitative argument to be the same: the $S_{(4)}$ signal should in general have higher visibility than $S_{(2)}$, and we expect there to be a considerable range of $J$ in which the cross peak is hidden in $S_{(2)}$ but visible in $S_{(4)}$.

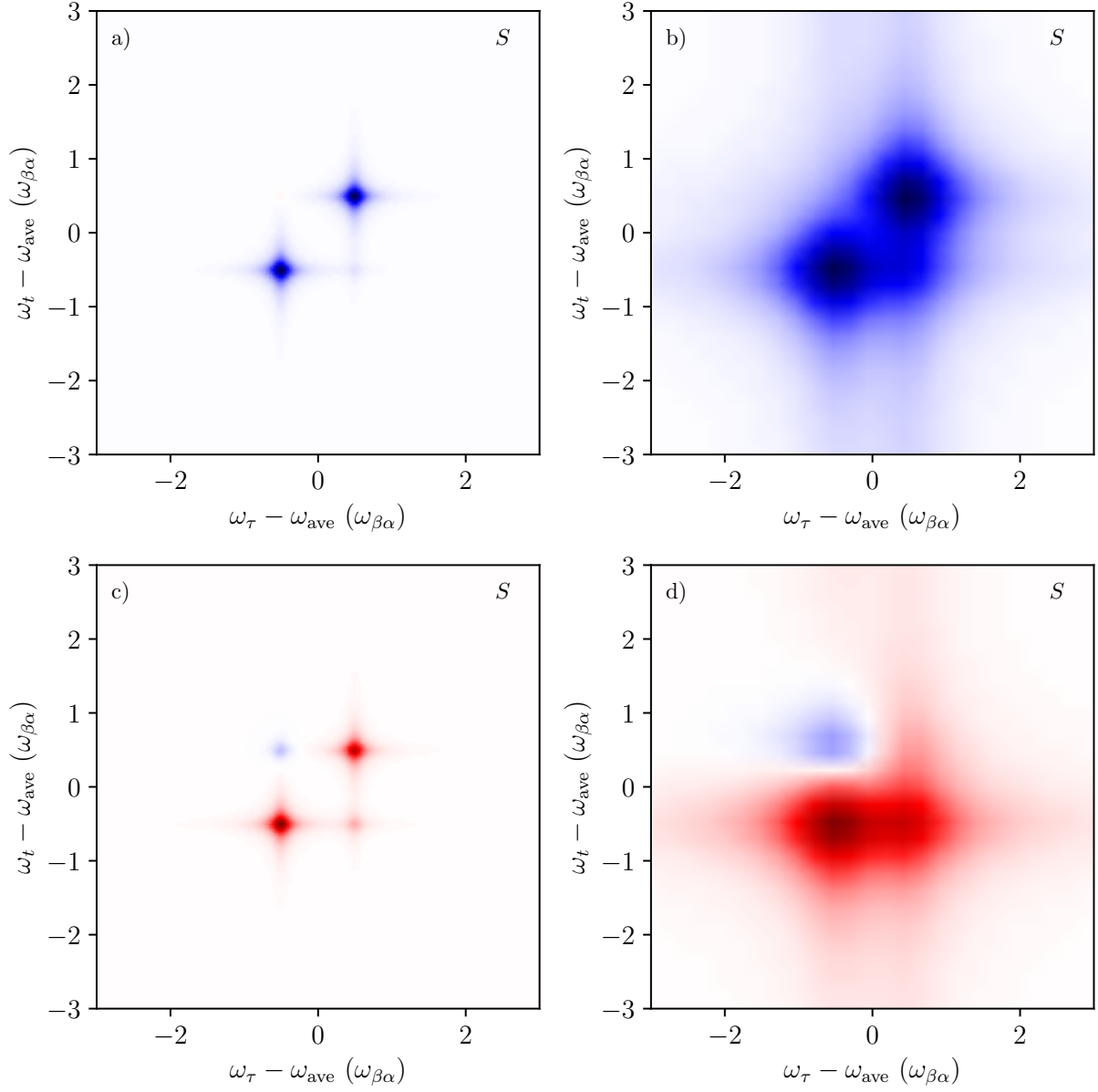

Figure S7: 1Q signals for the dimer model with $\mu_r = 1$, showing two different examples with no cross peak at position $(\omega_\tau, \omega_t) = (\omega_\alpha, \omega_\beta)$ visible at lowest order but clearly visible in $S_{(4)}$. (a) $S_{(2)}^{1Q}$ for $\Gamma = 0.1\omega_{\beta\alpha}$ and $J = 0.017\omega_{\beta\alpha}$. (b) $S_{(2)}^{1Q}$ for $\Gamma = 0.5\omega_{\beta\alpha}$ and $J = 0.1\omega_{\beta\alpha}$. (c) $S_{(4)}^{1Q}$ for $\Gamma = 0.1\omega_{\beta\alpha}$ and $J = 0.017\omega_{\beta\alpha}$. (d) $S_{(4)}^{1Q}$ for $\Gamma = 0.5\omega_{\beta\alpha}$ and $J = 0.1\omega_{\beta\alpha}$.

# References

(1) Malý, P.; Lüttig, J.; Turkin, A.; Dostál, J.; Lambert, C.; Brixner, T. From wavelike to sub-diffusive motion: exciton dynamics and interaction in squaraine copolymers of varying length. *Chem. Sci.* **2020**, *11*, 456–466.

(2) Völker, S. F.; Schmiedel, A.; Holzapfel, M.; Renziehausen, K.; Engel, V.; Lambert, C. Singlet-Singlet Exciton Annihilation in an Exciton-Coupled Squaraine-Squaraine Copolymer: A Model toward Hetero-J-Aggregates. *J. Phys. Chem. C* **2014**, *118*, 17467–17482.

(3) Lambert, C.; Koch, F.; Völker, S. F.; Schmiedel, A.; Holzapfel, M.; Humeniuk, A.; Röhr, M. I. S.; Mitric, R.; Brixner, T. Energy transfer between squaraine polymer sections: From helix to zigzag and all the way back. *J. Am. Chem. Soc.* **2015**, *137*, 7851–7861.

(4) Bubilaitis, V.; Abramavicius, D. Signatures of exciton–exciton annihilation in 2DES spectra including up to six-wave mixing processes. *J. Chem. Phys.* **2024**, *161*, 104106.

(5) Rose, P. A.; Krich, J. J. Efficient numerical method for predicting nonlinear optical spectroscopies of open systems. *J. Chem. Phys.* **2021**, *154*, 034108.

(6) Rose, P. A.; Krich, J. J. Automatic Feynman diagram generation for nonlinear optical spectroscopies and application to fifth-order spectroscopy with pulse overlaps. *J. Chem. Phys.* **2021**, *154*, 034109.



\end{document}